# Stress induced martensitic transformation in NiTi at elevated temperatures: martensite variant microstructures, recoverable strains and plastic strains

*O. Tyc, E. Iaparova, O. Molnárová, L. Heller, P. Šittner*

Institute of Physics of the Czech Academy of Sciences, Na Slovance 2, Prague, Czech Republic

**Abstract**

Functional behavior of NiTi shape memory alloys is restricted to temperatures below 150 °C. This is deemed to be partially due to relatively low transformation temperatures rendering high transformation stress and partially due to low resistance of the austenite and/or martensite phases to the plastic deformation at elevated temperatures and stresses. To shed light on the origin of the loss of functional properties of NiTi with increasing temperature, we have investigated stress induced martensitic transformations in nanocrystalline NiTi shape memory wire by thermomechanical testing supplemented with post-mortem reconstruction of martensite variant microstructures in grains by transmission electron microscopy.

It was found that the forward stress induced martensitic transformation proceeding via propagation of macroscopic Luders band fronts in stress plateaus is not completed at the end of the stress plateau and generates unrecoverable plastic strain in addition to recoverable transformation strain. The higher the test temperature, the larger is the plastic strain as well as the volume fraction of retained austenite. The reconstructed martensite variant microstructures in grains of the NiTi wire deformed up to the end of the stress plateau at 120 °C were found to be filled with single domains of (001) compound twin laminate, some grains were nearly detwinned and some grains contained (100) deformation twins. The recoverable transformation strain (~6%) generated by the forward martensitic transformation is nearly independent on the test temperature (plateau stress) because detwinning of (001) compound twins in the microstructure is prohibited by lateral constraint from the surrounding grains in nanocrystalline wire.

It is concluded that the gradual loss of functionality of NiTi with increasing temperature does not originate from the plastic deformation of austenite, as widely assumed in the literature, but that it derives from the lack of resistance of the stress induced martensite to the plastic deformation under increasing stress.

# 1. Introduction

The cyclic superelastic and actuation functionalities of NiTi shape memory alloy due to martensitic transformation (MT) from the B2 cubic austenite to B19' monoclinic martensite [1] are restricted to temperatures below 150 °C. This is partially due to the relatively low transformation temperatures rendering high transformation stress and partially due to low resistance of the austenite and/or martensite to the plastic deformation [2,3]. If NiTi alloys are strengthened via cold work/annealing and/or via introduction of $Ti_3Ni_4$ nanoprecipites into the austenitic microstructure, plastic strains generated during the the martensitic transformation can be suppressed so that they become small and accumulate for thousands of cycles. Although the strengthened nanocrystalline NiTi wires display superelasticity and shape memory effects at temperatures as high as ~150 °C [4], achieving satisfactory cyclic stability of stress-strain-temperature response of NiTi at elevated temperatures and stresses is very difficult. To be able to suppress the generation of plastic strains by manipulating the virgin austenitic microstructure of the alloy without affecting the recoverable transformation strains, we need to understand the mechanisms, by which forward and reverse MTs proceeding under stress generate plastic strains [3]

Based on the results of transmission electron microscopy (TEM) observation of dislocation defects in austenite generated by the stress induced martensitic transformation, Norfleet et al. [5], suggested that the plastic deformation during superelastic cycling proceeds via dislocation slip in austenite initiated at the habit plane interface. Researchers, who studied the forward stress induced MT in NiTi proceeding under external stress [6,7,8], assumed that the stress induced martensite created by the martensitic transformation is type II twinned, as predicted by the Phenomenological Theory of Martensite Crystallography (PTMC) [9]. Habit plane interface, as an undistorted and unrotated crystal plane, can be predicted from lattice parameters, lattice correspondence and lattice invariant shear (LIS) in martensite using the PTMC theory. Strain compatibility at the habit plane is fulfilled during the forward and reverse MTs under stress into type II twinned martensite. Slip dislocations observed in thermomechanically cycled NiTi [6,7,8,10] were assumed to nucleate at the propagating habit plane interface. Plastic strain generated by the slip of these dislocations during the forward MT, however, shall not be directly related to the strain compatibility at the habit plane. Moreover, plastic deformation via dislocation slip in austenite during the forward MT upon loading and-or cooling cannot assist the strain compatibility at the habit plane during the forward loading (plastic strain due to slip in austenite prior the moving habit plane interface is inherited by the martensite). This was briefly noted by Norfleet et al. [5] and later further elaborated by Sittner et al. [11]. On the other hand, plastic deformation of austenite behind the habit plane propagating during the reverse MT on unloading and-or heating may assist the strain compatibility at the habit plane.

The unrecovered plastic strain and lattice defects generated by the stress induced martensitic transformation were investigated by analyzing lattice defects in austenite using superelastic NiTi wire by TEM [5-8]. Since both forward and reverse MTs proceed under stress in superelastic cycle, it is not clear whether the lattice defects observed after a single cycle were created during the forward or reverse MT, whether they were created in the austenite or in the martensite phase. Indeed, the observed lattice defects could have been created by plastic deformation of martensite and only inherited by the austenite phase on unloading and/or heating. We have recently developed a thermomechanical testing method [2,3] enabling to investigate plastic strains and lattice defects generated by a single reverse and single forward MT taking place under stress. This method will be employed in this work to determine plastic strains generated by the forward stress induced MT in isothermal tensile tests on NiTi SME wire..

We employ two experimental methods for: i) reconstruction of martensite variant microstructures in whole grains of deformed NiTi by TEM [12,13,14] and ii) in-situ analysis of austenite and martensite textures evolving during tensile test on NiTi shape memory wires up to the fracture by in-situ x-ray diffraction [15]. We have already successfully applied these methods to reveal the deformation mechanisms of martensite reorientation [12,15] and plastic deformation of B19' martensite [16,13,15].

Recently, we started to apply these methods to reveal the mechanism by which the MT proceeding under stress generates plastic strains and lattice defects. In particular, we performed thermomechanical loading experiments on NiTi wires and evaluated martensite variant microstructures created by MT proceeding upon cooling under stress [17,18]. Preliminary results suggest that the B2-B19' martensitic transformation under stresses proceeds into (001) compound twinned martensite. The martensite created by stress induced martensitic transformation in tensile test on NiTi SME wire at 100 °C was found to exist as a partially detwinned single laminate of (001) compound twins filling whole grains of the deformed wire (Fig. 9 in [12]). However, the mechanism by which the forward MT generates plastice strains remained unclear.

As already mentioned, consideration of the stress induced martensitic transformation in NiTi in the literature [1,5-8] was significantly affected by the assumption on MT proceeding into type II twinned B19' martensite. Such MT was indeed frequently observed in NiTi single crystals and coarse grain polycrystals [1,19,20]. Matsumoto et al. [19] reported habit plane (-0.8684, 0.26878, 0.4138) of stress induced martensite in NiTi single crystal to be very close to theoretical PTMC solution for <011> type-II twinning as lattice invariant shear. Since this result concerned habit plane interface of the stress induced martensite, it has become generally accepted in the SMA field that the stress induced B19' martensite in NiTi forms by propagation of the habit plane interface giving rise to <011> type-II twinned habit plane variant in the martensite variant microstructure.

However, there is also growing experimental evidence that the stress induced martensite in NiTi is (001) compound twinned and/or detwinned [12-14,18,20,21,22,23]. However, since (001) compound twinning cannot serve as lattice invariant shear in the MT [1,9], it is not obvious how strain compatibility at habit plane interfaces is achieved. We have proposed [18] that the forward MT upon cooling NiTi wire under constant stress takes place via propagation of the habit plane between austenite and second order laminate of (001) compound twins, which immediately reorients into single laminate filling whole grains. Cayron [24] proposed that strain compatibility at the habit plane interface between austenite and detwinned martensite can be achieved taking into account plastic strain due to dislocation slip in austenite, modified correspondingly the PTMC theory and used it to explain the habit planes observed in in-situ EBSD studies of NiTi deformation [25]. Heller and Sittner [26] proposed a theoretical framework suggesting that strain compatibility at the habit plane interface of stress induced MT in NiTi can be achieved with the assistance of elastic strains in both austenite and martensite phases, though it generally requires high external stress. The calculated habit plane interface is thus not firmly given by lattice parameters and lattice correspondence but varies with the magnitude and orientation of the applied stress, which explains the large spread of experimentally determined habit planes between austenite and detwinned martensite (Fig. 6 in [27]).

As concerns the mechanism by which the forward martensitic transformation under stress generates plastic strains, key questions are whether the plastic deformation occurs in the austenite phase in front of the propagating habit plane or in the martensite phase behind it and how it is generated. Based on the results of experiments focusing forward MT upon cooling under constant stress [18], we now assume that it occurs [100](001) dislocation slip in martensite, which can be activated at relatively low applied stresses [28,29,30] but reliable experimental evidence is missing. When the B19' martensite is exposed to high stresses, it deforms plastically by kwinking [12,13,14,16] taking place via [100](001) dislocation based kinking combined with (100) deformation twinning. Nevertheless, mechanism of the forward stress induced martensitic transformation immediately followed by plastic deformation of martensite via dislocation slip and/or kwinking remains unclear - experimental data as well as theoretical concepts for such coupled forward martensitic transformation and plastic deformation are not available.

The aim of this work is to generate such experimental data and reveal the mechanism by which the forward martensitic transformation under stress generates plastic strains. To achieve that goal, recoverable transformation strains and unrecoverable plastic strains generated by forward MT in nanocrystalline NiTi wire in tensile deformation at various temperatures are evaluated using closed loop thermomechanical loading experiments and martensite variant microstructures created by the stress induced MT are reconstructed by nanoscale orientation mapping in TEM.

## 2. Experiment

NiTi shape memory wire produced by Fort Wayne Metals in cold work state (FWM #5 Ti-50.5 at. % Ni, 42 % CW, diameter 0.1 mm) was heat treated by short pulse of electric current [4] (power density 160 W/mm3, pulse time 15 ms). While performing the heat treatment, 30 mm long segment of cold worked wire was crimped by two steel capillaries, prestressed to ~300 MPa, constrained in length and subjected to 15 ms pulse of controlled electric power. The heat treated wire has a fully recrystallized microstructure with a mean grain size d = 250 nm (Fig.1). The heat treated alloy undergoes B2-R-B19' transformation with characteristic transformation temperatures $M_s$ = 63 °C, $A_f$ = 93 °C upon cooling/heating, as evaluated combined in-situ electric resistivity and dilatometry tests (Fig.1). The 15 ms NiTi #5 wire was selected because of its recrystallized nanoscale austenitic microstructure free of lattice defects from previous cold work required for post-mortem TEM investigations, because it is martensitic at room temperature and because the stress induced MT is accompanied by generation of large plastic strains even if it takes place at low applied stresses [17].

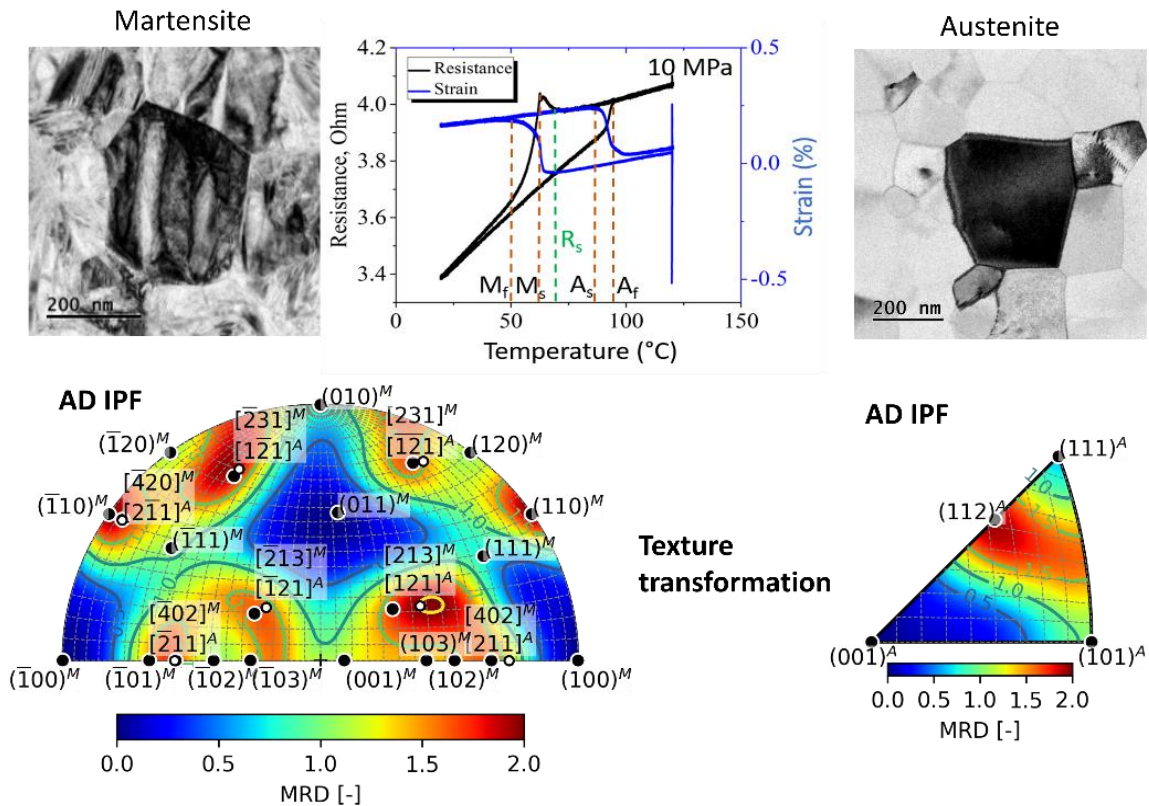

**Figure 1: Thermally induced B2-B19' martensitic transformation in 15 ms NiTi #5 shape memory wire.** Transformation temperatures evaluated by combination of dilatometry and electric resistivity measurement during thermal cycle under 10 MPa tensile stress. TEM figures show fully recrystallized austenitic microstructure with mean grain size of 250 nm and self-accommodated martensitic microstructure consisting of multiple (001) compound twin laminates in a grain. The <111> fiber texture of austenite transforms upon cooling into four (8)-pole fiber texture of self-accommodated martensite [18].

Tensile tests were carried out using an in-house-built tensile tester (MITTER) consisting of a miniature load frame, an environmental chamber, electrically conductive grips, a load cell, a linear actuator and a position sensor. The environmental chamber (Peltier elements) enables to maintain homogeneous temperature around the thin wire from -30 °C to 200 °C. The wire sample was gripped into the tester, strain was set to zero in the austenitic state at 200 °C, the wire was cooled/heated under 20 MPa stress to the desired test temperature and tensile thermomechanical loading tests were performed. Thermomechanical loading tests were performed in combined position and force control mode (cooling rate of 3 °C/min). Thermomechanical tests involving cooling down to −120 °C and above 200 °C were performed using DMA 850 tester by TA Instruments using 10 mm long wire samples. Thermal strains were calibrated using quartz sample and subtracted from the signal recorded in thermomechanical tests.

Thin lamellae for TEM analysis were cut from the subsurface layers of deformed wire (10 μm below the surface, wire axis in the lamella plane) with the wire axis in the lamella plane by focused ion beam (FIB) using a FEI Quanta 3D FIB-SEM microscope. Martensite variant microstructures and interfaces in grains of deformed NiTi wire were analysed by TEM using a FEI Tecnai TF20 X-twin microscope equipped with a field emission gun operating at 200 keV using a double tilt specimen holder. The recorded electron diffraction patterns were indexed using the lattice parameters of the B2 and B19' structures ($a_0$ = 0.3015nm, a=0.2889, b=0.4120, c=0.4622, β=96.8)

Martensite variant microstructures in grains of plastically deformed NiTi wire were analysed by Selected Area Electron Diffraction with Dark Field (SAED-DF) method [12,1314] and by the automated orientation mapping ASTAR method [31]. Reconstruction of the martensite variant microstructure by the ASTAR method is achieved by scanning the selected grain by precessing electron beam while recording electron diffraction patterns and images by fast CCD camera [31]. For more information on the ASTAR reconstruction of martensite variant microstructure in NiTi see Ref. [14].

When using the SAED-DF and ASTAR methods to reconstruct the martensite variant microstructures in whole grain [14], TEM lamella is tilted in such a way that the selected grain is oriented into <010> low index zone (<010> denotes [010] and/or [0-10] direction of the monoclinic lattice). The <010> zone is selected because the martensite variants in grains arrange themselves in such a way that all crystal lattices within a single grain are equally oriented with respect to the electron beam in the [010] zone and all interface planes are parallel to the electron beam [14]. This is direct consequence of the activated deformation mechanisms, particularly to martensite reorientation and plastic deformation of martensite by kwinking [16].

## 3. Results

To evaluate recoverable and plastic strains generated by the forward MT under external stress, we need to know the [temperature, stress] conditions, at which the forward MTs take place and critical yield stresses at which the austenite and martensite start to deform plastically. Fig. 2e shows stress-temperature diagram (σ-T diagram) presenting critical [temperature, stress] conditions for activation of various deformation/transformation processes in the 15ms NiTi #5 wire loaded in tension. The diagram was constructed from the results of isothermal and isostress tests in temperature range -120° C - 550 °C [32]. Figs. 2b,c,d show selected stress-strain curves of 15 ms NiTi #5 wire deformed in isothermal tensile tests until fracture in a wide temperature range 20° C - 150 °C.

The stress-strain curves recorded at ambient temperatures (Figs. 2d) display two stress plateaus. In the first stress plateau, the wire deforms via reorientation in the martensite state ($\sigma^{RE}$) at low temperatures (< 60 °C) and via stress induced martensitic transformation ($\sigma^{FOR}$) at high temperatures (> 90 °C). The stress plateaus reflect the localized deformation of the wire via propagation of macroscopic Luders band fronts either during the martensite reorientation or during the stress induced MT [33]. In the second plateau, the wire deforms plastically via kwinking deformation [16]. Between both stress plateaus, the wire shows linear stress-strain response, the slope of which decreases with increasing temperature (Fig. 2d).

While plateau stress increases, kwinking stress decreases with increasing test temperature (Fig. 2e), as a result of which the forward MT and kwinking lines in the σ-T diagram meet at ~800 MPa stress and 150 °C. Theoretically, NiTi wire deformed in tensile tests at elevated temperatures 80-150 °C shall undergo stress induced MT in the plateau range followed by elastic deformation of oriented martensite beyond the end of the plateau and plastic deformation beyond the yield stress for kwinking. In reality, plastic deformation accompanies the forward MT in the plateau range as well as during tensile loading of the wire beyond the end of the plateau (Fig. 2a).

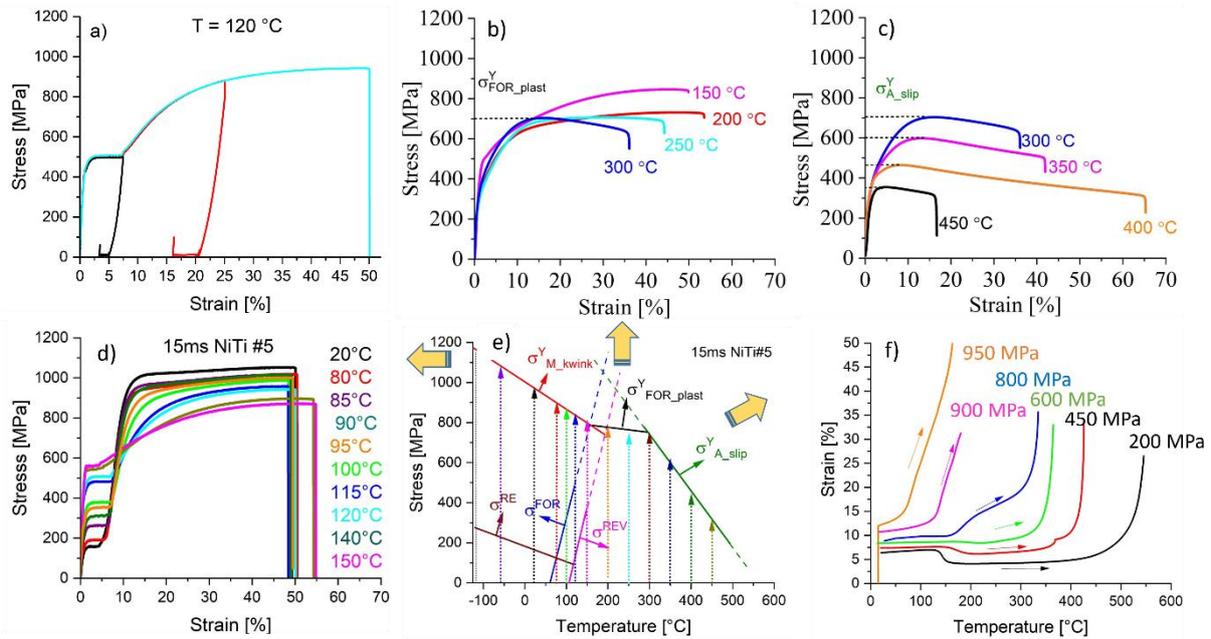

**Figure 2: Stress-strain-temperature curves (a,b,c,d,f) determined in tensile thermomechanical loading tests on 15 ms NiTi #5 shape memory wire**. a) tensile loading at 120 °C up to fracture, end of the stress plateau and 25% strain followed by unloading and heating up to 200°C, b,c,d) stress-strain curves from tensile test until rupture in different temperature ranges, f) strain-temperature curves from thermomechanical loading tests involving heating wires deformed in martensite state up to various stresses and heating under constant stress till rupture [32] and e) stress-temperature diagram showing critical [temperature, stress] conditions for activation of various deformation/transformation processes in the wire determined from thermomechanical loading tests (b,c,d,f) and stress-temperature path of tensile tests at constant temperature denoted by arrows.

In tensile tests at test temperatures higher than 150 °C (Figs. 2b,c), there are no stress plateaus on stress-strain curves anymore, the alloys displays remarkable strain hardening at high stresses around 900 MPa. According to the deformation mechanism responsible for the stress-strain response, there are two different regimes at high temperatures. In the temperature range 150 °C - 300 °C (Fig. 2b), the wire deforms via stress induced MT coupled with plastic deformation ($\sigma^Y_{FOR\_plast}$). The capacity for strain hardening, however, decreases with increasing test temperature, as a result of which, the ductility decreases. In the temperature range 300 °C - 450 °C (Fig. 2c), the wire deforms plastically via dislocation slip in austenite ($\sigma^Y_{A\_slip}$). This deformation mode is characterized by strain softening, strain rate sensitivity and decreasing flow stress with increasing temperature. The ductility increases with increasing temperature. When the wire is heated under constant stress (Fig. 2f), it undergoes reverse MT and deforms via dislocation slip in austenite at low stresses 200 MPa and via kwinking deformation at high stress above 900 MPa [32]. Upon heating under 800 MPa stress, all three deformation processes take place [32].

Within the "generalized elasticity area" below the red $\sigma^Y_{M\_kwink}$ line, black $\sigma^Y_{FOR\_plast}$ and green $\sigma^Y_{A\_slip}$ line in the σ-T diagram (Fig. 2e), the 15 ms NiTi #5 shape memory wire shall deform only elastically and via deformation processes derived from MT. Nevertheless, as mentioned in the introduction, it undergoes incremental plastic straining whenever the forward and/or reverse MTs take place under external stress (Fig. 2a). We are particularly interested in the mechanism by which forward stress induced martensitic transformation in the plateau range generates incremental plastic strains. As the test temperature increases in the temperature range 70-150 °C, plateau stress increases, kwinking stress decreases, slope of the stress-strain curve beyond the end of plateau decreases (Fig. 2d). Although all recorded stress-strain curves in Fig. 2d appear to be similar, the activated deformation mechanism dramatically changes with increasing temperature. While the wire deformed close above $M_s$ temperature at 75°C shows almost perfect strain recovery of upon heating, the wire deformed at 150 °C shows very small recoverable strain in tensile tests up to similar maximum strains ~5%. This was investigated in thermomechanical loading tests focusing tensile deformation up to the end of the stress plateau followed by strain recovery upon unloading and heating (Fig. 3). The recoverable transformation strains $\varepsilon^{tr,rec}$ increases up to 100 °C, but then decrease with increasing test temperature (Fig. 3c).

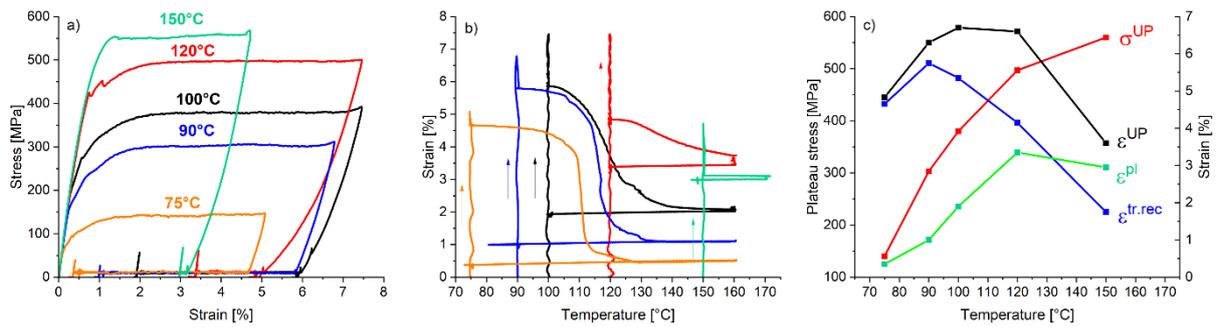

**Figure 3: Results of thermomechanical loading tests on 15 ms NiTi #5 shape memory wire involving tensile deformation at various temperatures followed by heating to 160C under 20 MPa.** a) stress-strain curves, b) strain-temperature curves, c) temperature dependence of plateau strain $\varepsilon^{UP}$, recoverable transformation strain $\varepsilon^{tr,rec}$, plastic strain $\varepsilon^{pl}$ and plateau stress $\sigma^{UP}$,

But why is that? This is not very clear. The loss of the strain recoverability with increasing test temperature [34,35] is usually ascribed in the SMA literature to the onset of plastic deformation of austenite at $M_D$ temperature [35]. In some sense, this is true because the wire eventually starts to deform via dislocation slip in austenite, though, according to figure 2, it only occurs in tensile tests at test temperatures above 300 °C (Fig. 2c). The $M_D$ temperature is thus somewhere between 250 and 300°C. However, there is a wide transition range of test temperatures 70-300 °C (Figs. 2b,d), in which the 15 ms NiTi #5 wire the stress induced martensitic transformation is accompanied by plastic deformation, the character of which gradually changes with increasing temperature. In our opinion, better understanding of the deformation mechanism in this temperature range is a key to the understanding of high temperature limits and loss of functionality of NiTi with increasing temperature.

Experimental observations from tensile tests, which evidence this gradual change of deformation mechanism with increasing temperature, are: i) loss of the strain recoverability (Fig. 3), ii) decrease of the slope of the stress-strain curve beyond the end of the stress plateau (Fig. 4a) and iii) temperature dependence of plateau stress deviating from linearity (Fig. 4a).

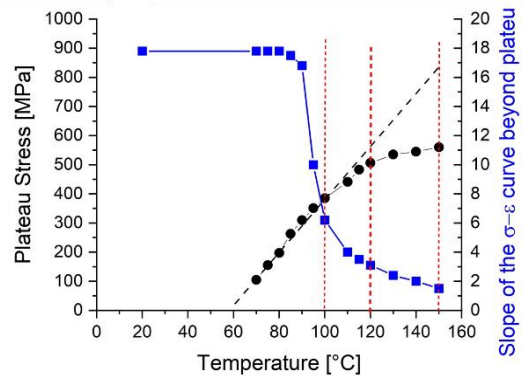

**Figure 4: Additional results of the analysis of tensile tests at temperatures 20-150 °C (Fig. 2d):** plateau stress and slope of the σ–ε curve beyond the end of the stress plateau plot in dependence on the test temperature,

The slope of the stress-strain curve in this stage (Fig. 4a) stays constant with temperature increasing from -90 °C up to 80 °C suggesting that the stress-strain response beyond the end of the stress plateau at low temperatures is temperature independent. Whether the stress response is true linear elastic remains questionable, but it is recoverable on loading-unloading. But when the test temperature increases above 80 °C, the slope sharply decreases (Figs. 2a) and the stress response becomes unrecoverable on loading-unloading. This means that stress-strain response is not elastic anymore at these temperatures. The decrease of the slope with increasing temperature is either due to the stress induced martensite deforming plastically and/or due to the retained austenite transforming to martensite upon tensile loading.

The temperature dependence of plateau stress shall follow the linear relationship described by the Clausius–Clapeyron equation [1]. Nevertheless, this is not the case in real experiments (Fig. 4a). The temperature dependence of plateau stress deviates from the linearity in the same temperature range, where the slope of the stress-strain curve decreases (Fig. 4a) and shape memory becomes gradually lost (Fig. 3). The stress-strain curve recorded in the tensile test at 120 °C, unloading and heating up to 200 °C (Fig. 2a) serves as an example of material behavior which combines stress induced martensitic transformation with plastic deformation. The deformation mechanism activated in this test will be investigated in detail in section 3.2.

**3.1 Stress induced martensitic transformation in tensile tests at elevated temperatures**

The key to the understanding of the mechanism by which the NiTi wire deforms via stress induced martensitic transformation is to know the recoverable transformation strain and plastic strains generated by the forward MT on loading. The thermomechanical loading test in Fig. 3 is not proper experimental way to investigate the forward MT proceeding under stress because the plastic strain potentially generated by the reverse MT on unloading makes it impossible to evaluate the recoverable transformation strain and plastic strains generated by the forward MT. Therefore, we performed slightly different thermomechanical loading tests on the 15 ms NiTi #5 wire presented in Fig. 5.

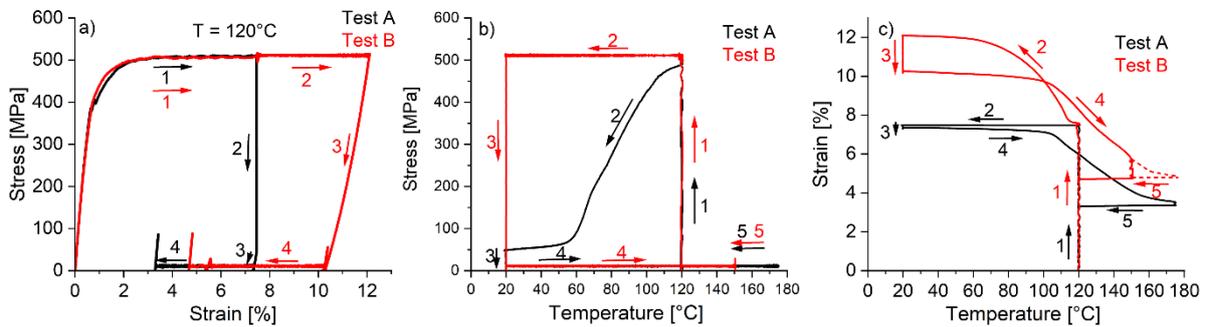

**Figure 5: Stress-strain-temperature responses of 15 ms NiTi #5 shape memory wire in two thermomechanical loading tests featuring stress induced MT at 120 °C (a,b,c)**. The experiments were designed to verify a suspicion that the wire deforms plastically in the plateau range and that the stress induced MT is not completed at the end of the plateau, The wire was deformed to the end of the stress plateau at 120 °C, and then it was either cooled to room temperature under constant length, then unloaded and heated under 20MPa stress up to 200 °C (Test A) or it was cooled under 500 MPa plateau stress to room temperature, unloaded and heated under 20MPa stress up to 200 °C to evaluate plastic strain (Test B). The strain-temperature response recorded in the test B confirms that there was indeed a retained austenite at the end of plateau stress, which transformed to oriented martensite upon cooling (stage 2). Both recoverable and unrecovered plastic strains are larger in the test B than in the test A.

To evaluate recoverable transformation strain and plastic strains generated by the forward MT under stress only [2,3], we need to perform closed loop thermomechanical loading tests, in which the reverse MT proceeds in the absence of stress (i.e. the unloading at high temperatures during which plastic strain can be generated has to be avoided). Thermomechanical loading test was hence performed as introduced in Fig. 5. The wire was deformed up to the end of the stress plateau, cooled to room temperature under constant strain, unloaded and heated above the $A_f$ temperature in the test A (black curves in Fig. 5). The stress gradually decreases to ~50 MPa upon cooling to Ms temperature under constant strain (stage 2 in Fig. 5b). The recovered and unrecovered strains in the close loop thermomechanical cycle were evaluated from the stress-strain-temperature response. Alternatively, the wire was deformed up to the end of the stress plateau, but then it was cooled under constant applied stress to the room temperature, unloaded and heated above the $A_f$ temperature in the test B (blue curves in Fig. 5). The strain further increased during the cooling under constant stress (stage 2 in Fig. 5c) because there was retained austenite in the wire, which transformed to oriented martensite upon cooling under constant stress in the test B. The forward MT continues during the cooling in both tests A and B, the difference is in the condition under which the cooling occurs. The advantage of test B is that the forward MT is completed and that we can evaluate cooling strain which is linked to the amount of retained austenite. Both recoverable and unrecovered plastic strains are generated by the forward MT in stages 1 and 2. No plastic strains are generated by the reverse MT upon heating in the absence of stress.

Results of a set of thermomechanical loading experiments (test B), in which stress induced forward MT proceeds at different temperatures, are shown in Fig. 6. The wire was subjected to tensile thermomechanical loading tests in which the forward stress induced MT took place at temperatures 70-150 °C. On the low temperature side at 70 °C, the transformation plateau stress is only ~100 MPa and strains generated on tensile loading are fully recoverable on unloading and heating. However, the higher the test temperature, the larger is the strain generated on loading within the plateau but also the cooling strain generated during the subsequent cooling under constant stress. On the upper temperature side at 150 °C, transformation plateau stress is ~560 MPa (Figs. 2a,6,7), plateau strain decreases to ~5%, very large strain ~12% is generated during cooling under constant stress and the unrecovered plastic strain reaches ~8% exceeding the recoverable transformation strain generated by the forward MT (Fig. 7). Figure 7 summarizes the change of the total inelastic strain, plateau strain, strain observed on cooling, recovered strain and unrecovered strain with increasing test temperature.

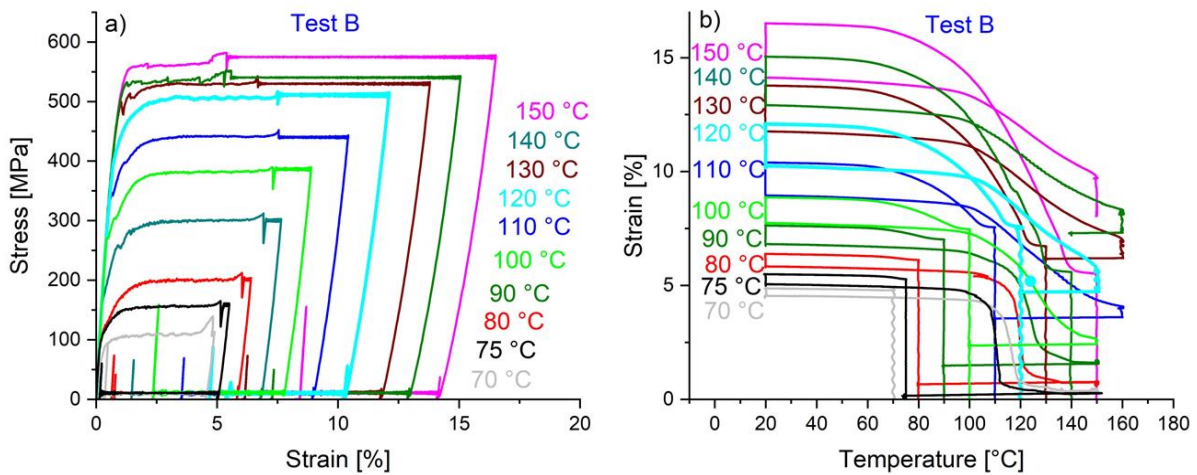

**Figure 6: Results of a set of thermomechanical loading tests on 15 ms NiTi #5 shape memory wire** involving tensile deformation up to the end of the stress plateau at constant temperatures 70-150 °C followed by cooling under constant plateau stress to room temperature and heating under 20 MPa up to 160 °C (test B in Fig. 5). a) stress-strain responses, b) strain-temperature responses.

Although the strain generated by the forward MT on tensile loading and cooling under stress continuously increases with increasing temperature (Fig. 6,7), the recoverable strain, after initial increase up to 80 °C (200 MPa), remains approximately constant (~5%) – i.e. does not depend on the test temperature and stress, at which the stress induced forward MT takes place. This is very different from the results of standard thermomechanical loading tests in Fig. 3. There are two reasons for the experimentally observed decree of recoverable strain ε$^{tr,rec}$ with increasing temperature: 1) the forward MT within the stress plateau in tests in Fig. 3 becomes less and less completed, 2) plastic deformation generated by the reverse MT on unloading compensates recoverable transformation strain [3].

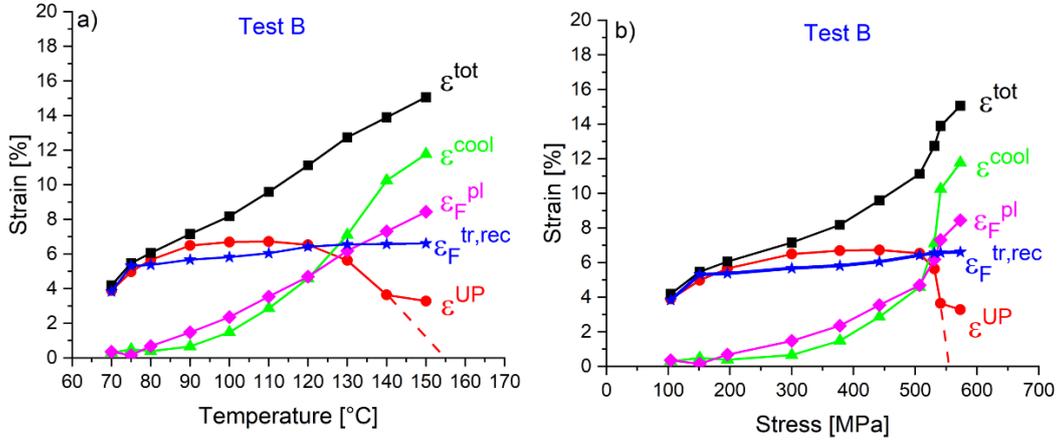

**Figure 7: Strain components evaluated from the stress-strain-temperature responses recorded in tensile thermomechanical loading tests in Fig. 6.** The total inelastic strain ε$^{tot}$, upper plateau strain ε$^{UP}$, cooling strain ε$^{cool}$, recoverable transformation strain ε$_F^{tr,rec}$ and unrecovered plastic strain ε$_F^{pl}$ generated by the forward MT under external stress are plotted in dependence on the test temperature (a) and plateau stress (b).

### 3.2 Stress induced martensitic transformation in tensile test at 120 °C

Let us analyze the deformation mechanism activated in the tensile test at 120 °C, in which the forward MT is not completed within the stress plateau. The stress-strain curve (Fig. 2a) displays elastic deformation of austenite, transformation stress plateau presumably due to the stress induced MT and quasielastic deformation of martensite upon tensile loading beyond the end of the stress plateau. Similar deformation behavior of NiTi alloys is frequently called superelasticity in the literature. However, this is questionable, since only less than half of the transformation plateau strain is recovered on unloading and heating. Within the plateau range, the wire deforms via a deformation mechanism which involves stress induced martensitic transformation coupled with plastic deformation. When loaded further beyond the end of the stress plateau, the wire does not deform only via elastic deformation, but the coupled MT with plastic deformation most likely continues, since the slope of the stress-strain curve is very low.

In order to investigate this coupled MT with plastic deformation in detail, the wire was deformed at 120 °C up to maximum strains 7.5%, 12%, 15% and 25%, cooled to room temperature, unloaded and heated to 120 °C (Fig. 8). The forward MT in these tests was again completed by cooling either under constant strain (test A) or under constant stress (test B). The results (total inelastic strain ε$^{tot}$, cooling strain ε$^{cool}$, recoverable transformation strain ε$_F^{tr,rec}$ and unrecovered plastic strain ε$_F^{pl}$ in Fig. 9b) show that the recoverable strain generated by the forward MT ε$_F^{tr,rec}$ remains constant but the cooling strain ε$^{cool}$ decreases with increasing maximum strain (stress). This implies that the forward MT continues while the wire is further loaded beyond the end of the stress plateau (the volume fraction of the retained austenite decreases) but generates mainly plastic strains (plastic strain increases and recoverable strain remains constant). The forward MT thus continues upon tensile deformation beyond the end of the stress plateau but generates mainly plastic strains (recoverable strain reached its limit while the wire was deformed within the stress plateau range).

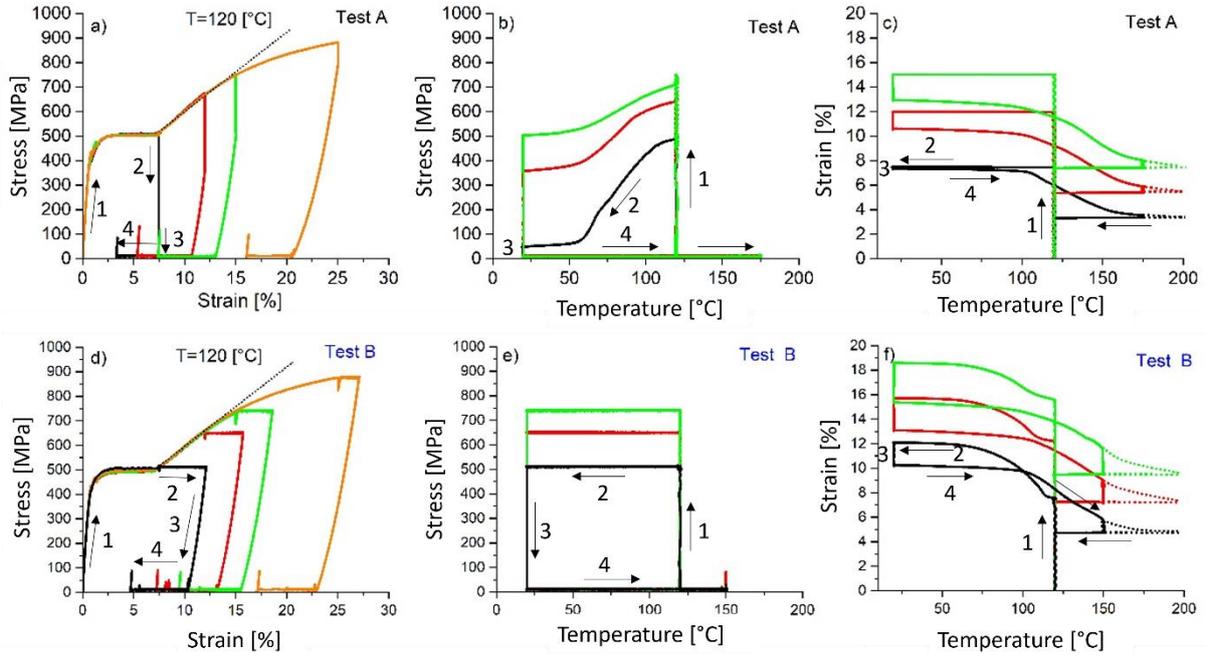

**Figure 8: Stress-strain-temperature responses of 15 ms NiTi #5 shape memory wire in four thermomechanical loading tests at 120 °C.** The wire was deformed up to increasing maximum strain at 120 °C and then, it was either cooled to room temperature under constant length, then unloaded and heated under 20MPa stress up to 200 °C (Test A in (a,b,c)) or it was cooled under 500 MPa stress to room temperature, then unloaded and heated under 20MPa stress up to 200 °C (Test B in (d,e,f)).

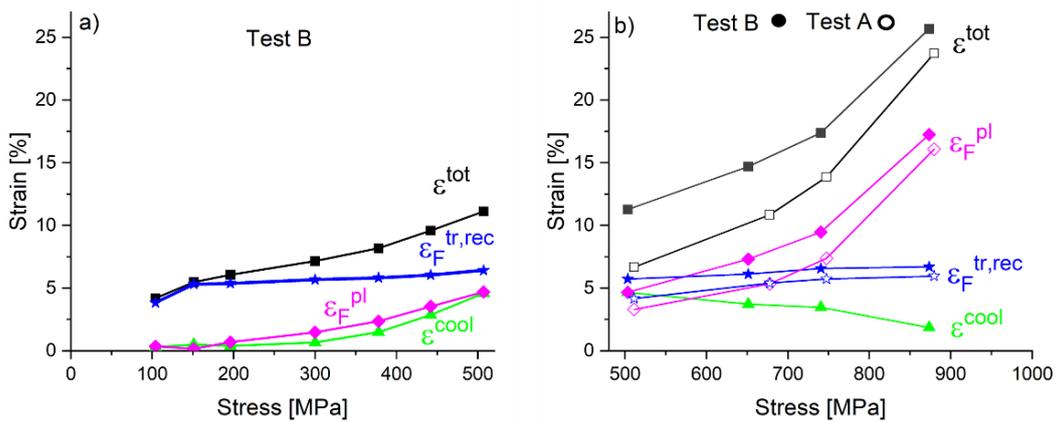

**Figure 9: Strain components evaluated from the stress-strain-temperature responses recorded in tensile thermomechanical loading tests.** The total inelastic strain $\varepsilon^{tot}$, cooling strain $\varepsilon^{cool}$, recoverable transformation strain $\varepsilon_F^{tr,rec}$ and unrecovered plastic strain $\varepsilon_F^{pl}$ generated by the forward MT under external stress are plot in dependence on the maximum applied stress. a) tensile deformation up to the end of the upper stress plateau at various temperatures (Fig. 6), b) tensile deformation at 120 °C beyond the end of the stress plateau up to increasing stress (Fig. 8). The strain components evaluated from tests B are systematically larger than strains evaluated in tests A.

## 3.3 Martensite variant microstructures in the NiTi wire deformed in tensile test at 120 °C

To investigate the mechanism by which the stress induced martensitic transformation in tensile test at 120 °C generates plastic strains, martensite variant microstructures in wires deformed up to the end of the plateau at 7.5% and further up to 12% and 15% strains were analyzed in TEM (Fig. 10). The arrangement of martensite variants in martensite variant microstructures is linked to the recoverable transformation strains generated by the forward MT. It can be also used to detect activity of plastic deformation of martensite by kwinking. However, it is very difficult to obtain information on plastic deformation by dislocation slip in martensite.

When analyzing martensite variant microstructures at room temperature, we assume that the observed microstructure existed in the wire under external stress. However, there is a serious problem with observation of martensite variant microstructures that existed in NiTi wires deformed in tensile test at high temperatures. If the wire is unloaded at high temperature and subsequently cooled to room temperature, the martensite would reverse transform to austenite on unloading and this austenite will transform to self-accommodated martensite upon stress-free cooling. To avoid that, the tests are finished by cooling to room temperature under constant strain (test A) or constant stress (test B) unload and cut the TEM lamella from the wire (Figs. 5,6,9). In this way, we can prepare TEM lamellae with oriented martensite even if there is significant fraction of retained austenite in the NiTi wire deformed at 120 °C (Figs. 8,10).

Bright field TEM images from the deformed wire (Fig. 10a,d,g) show polycrystal grains filled with deformation bands. However, it is not possible to obtain any meaningful information on martensite variant microstructures from the Fig. 10, since the microstructures in grains appear in general orientations with respect to the electron beam. Nevertheless, when the lamella is tilted, some grains turn completely dark. When we look into these grains (Fig. 10b-i) we can see characteristic martensite variant microstructures existing in the wire deformed up to various maximum strains. We learned that in our previous research involving analysis of martensite variant microstructures in grains of plastically deformed martensitic NiTi wire [12]. It was ridiculous, since it meant that all martensite variants within these grains diffract strongly, since they are oriented in low index zones. Nevertheless, it was found that this is exactly the case [12,14].

To obtain a useful information on the stress induced MT within the stress plateau and subsequent deformation beyond the end of the plateau (Fig. 8), we need to reconstruct the martensite variant microstructures in grains of NiTi wire deformed up to strain within this deformation range. We used the manual SAED-DF method [12,14] alongside the automated ASTAR method [14,31] for that purpose.

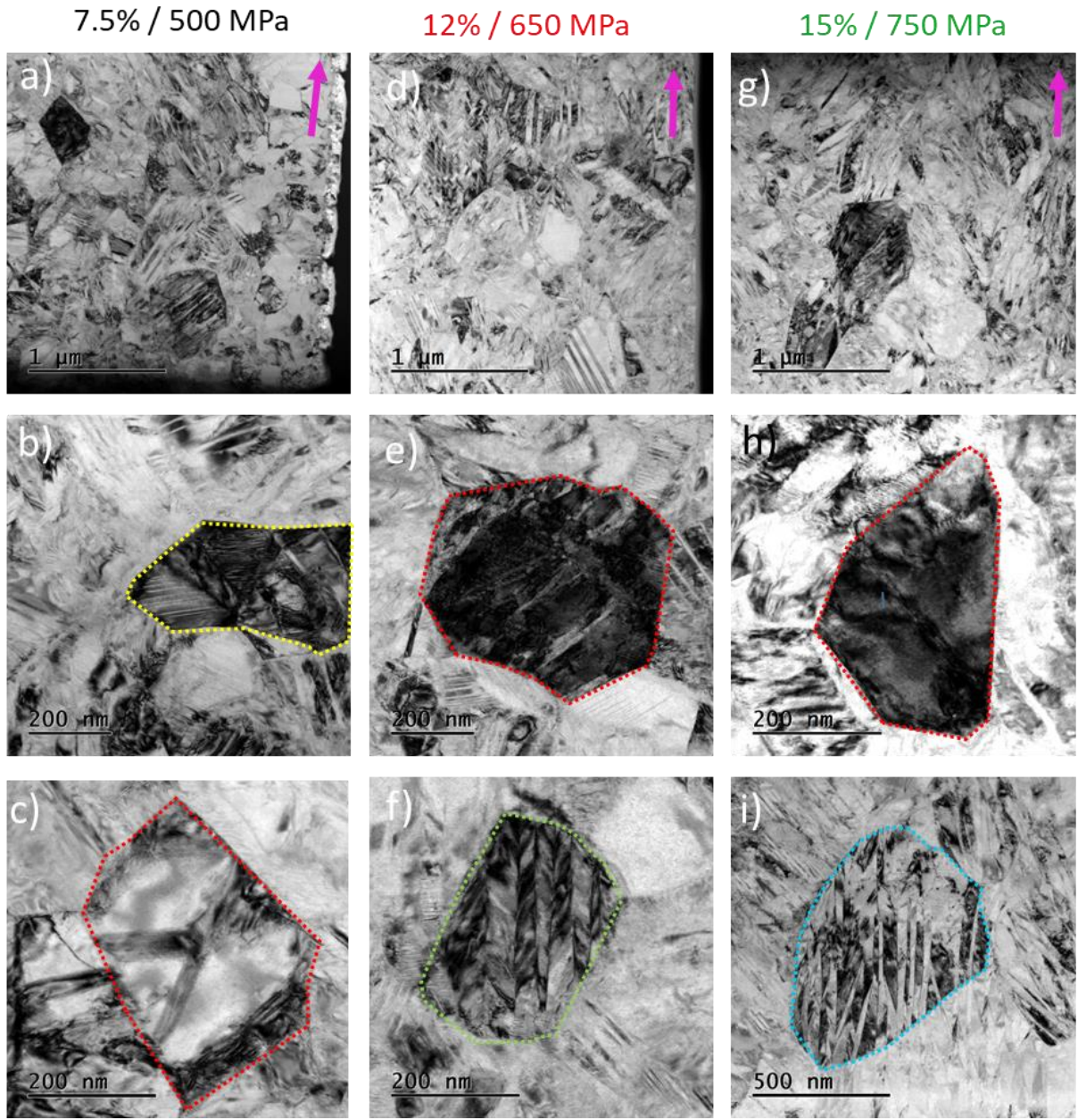

**Figure 10: Bright field TEM images of martensite variant microstructures existing in grains of 15 ms NiTi #5 shape memory wire** when loaded up to maximum strains in tensile tests in Fig. 8: a,b,c) 7.5% at the end of the stress plateau, d,e,f) 12% strain and g,h,i) 15% strain. Upper row shows low magnification overview of the martensite variant microstructures in the lamella, middle row details of selected grains oriented in low index zone and bottom row grains in general orientation in which typical martensite variant microstructures in grains can be observed. Grains with (001) compound twin laminate are decorated by yellow, detwinned grains by red, grains with (100) twins by green and grains containing microstructure of plastically deformed martensite with (20-1) kwink band by cyan colored grain boundaries.

The analysis of martensite variant microstructures in grains of the NiTi wire deformed up to the end of the stress plateau at 7.5% strain (Figs. 11-13) showed: i) grains filled with single domain (001) compound twin laminate (Fig. 11), partially detwinned grains (Fig. 12) and iii) partially detwinned grains containing few (100) deformation twins (Fig. 13). The martensite variant microstructures are reconstructed in a form of ASTAR datasets containing 3D spatial and orientation information on crystal lattices in individual martensite variants. It is essential to understand that we do not analyze a peculiarity. Any grain of the deformed polycrystalline wire contains a qualitatively similar microstructure that can be reconstructed by the ASTAR or SAED-DF method when properly oriented with respect to the electron beam [14].

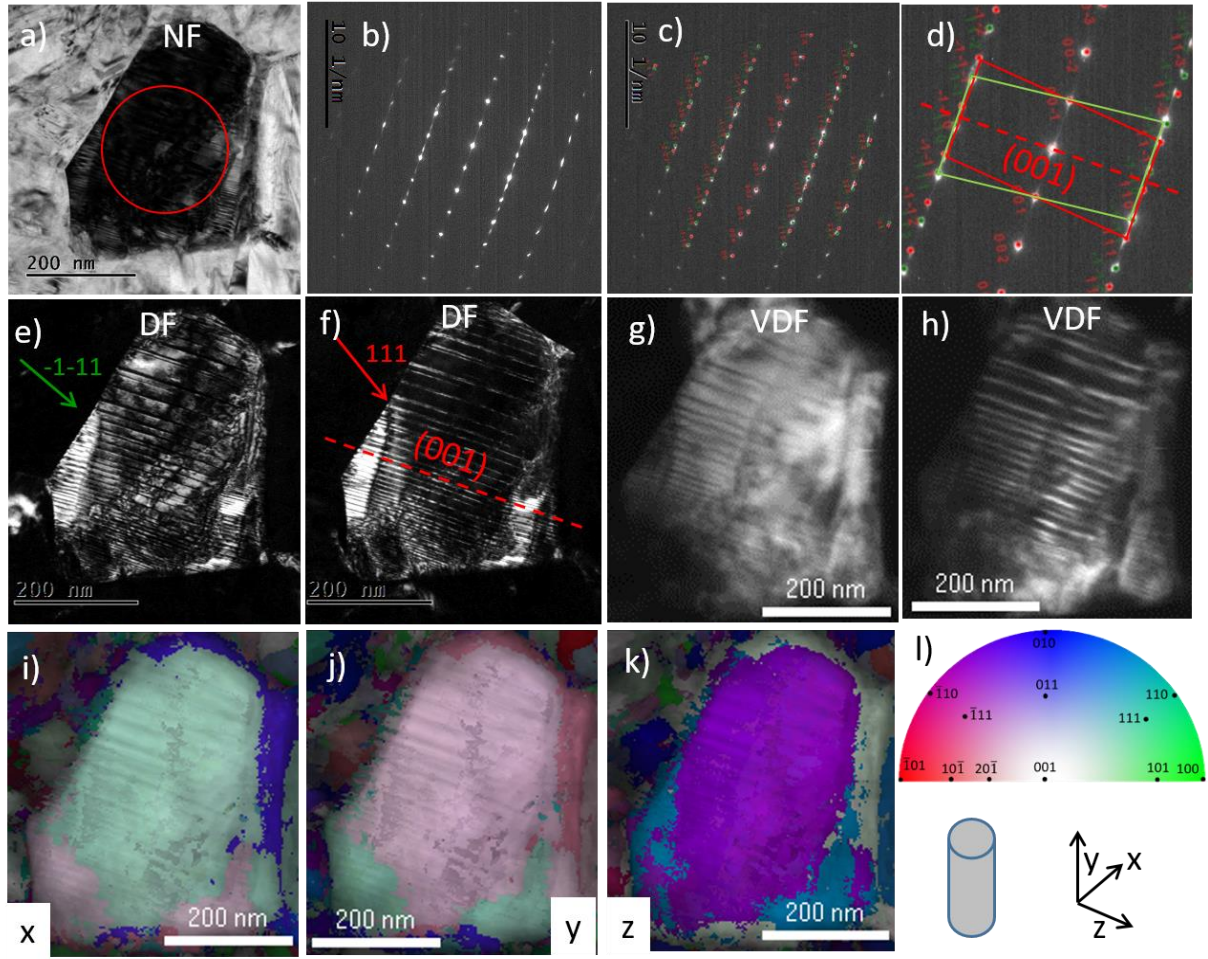

**Figure 11: Martensite variant microstructures in a grain** filled with a single laminate of (001) compound twins in 15 ms NiTi #5 shape memory wire deformed up to end of the stress plateau 7.5% strain (**Fig. 8**). a) BF image of the selected grain oriented in [1-10] zone, b,c,d) electron diffraction pattern taken from the SAED area denoted in (a) showing two overlapping diffraction patterns corresponding to B19' lattices in (001) compound twin relation, e,f) Dark field (DF) images of the grain taken using denoted reflections of both martensite lattices visualizing the (001) compound twin laminate filling whole grain, g,h) Virtual dark field (VDF) images from the analysis of the ASTAR map of the grain, i,j,k) ASTAR reconstruction of the martensite variant microstructure in the grain presented by showing crystal orientations within the grain along the x-,y- and z-axis (along the electron beam roughly perpendicular (α tilt 10°, β tilt -8.5°) to the lamella surface). Color scale in (l) denotes orientations of basic crystal directions of the monoclinic lattice.

We present the observed martensite variant microstructures in few selected grains (Figs. 11-15) by showing SEAD patterns, dark field (DF) images, color orientation maps along x-, y, and z- directions and virtual dark field (VDF) images evaluated from the ASTAR dataset representing the 3D reconstructed martensite variant microstructure in the selected grain. More information on the SAED-DF and ASTAR methods used to reconstruct martensite variant microstructure in grains of deformed NiTi wire can be found in Ref. [14].

The key result is that most of the grains of NiTi wire deformed up to 7.5% strain via stress induced MT within stress plateau at 500 MPa are filled with partially detwinned single laminate of (001) compound twins (Figs. 11,12) and few isolated (100) twin bands (Fig. 13). This was not surprising for us, since very similar microstructures were observed in the same wire deformed at 100 °C (Fig. 9 in [12]) which was assumed [12] to deform via stress induced MT. Besides of that, however, more than 3 % plastic strain was generated by the forward MT within the plateau range in tensile test at 120 °C (Figs. 3,8).

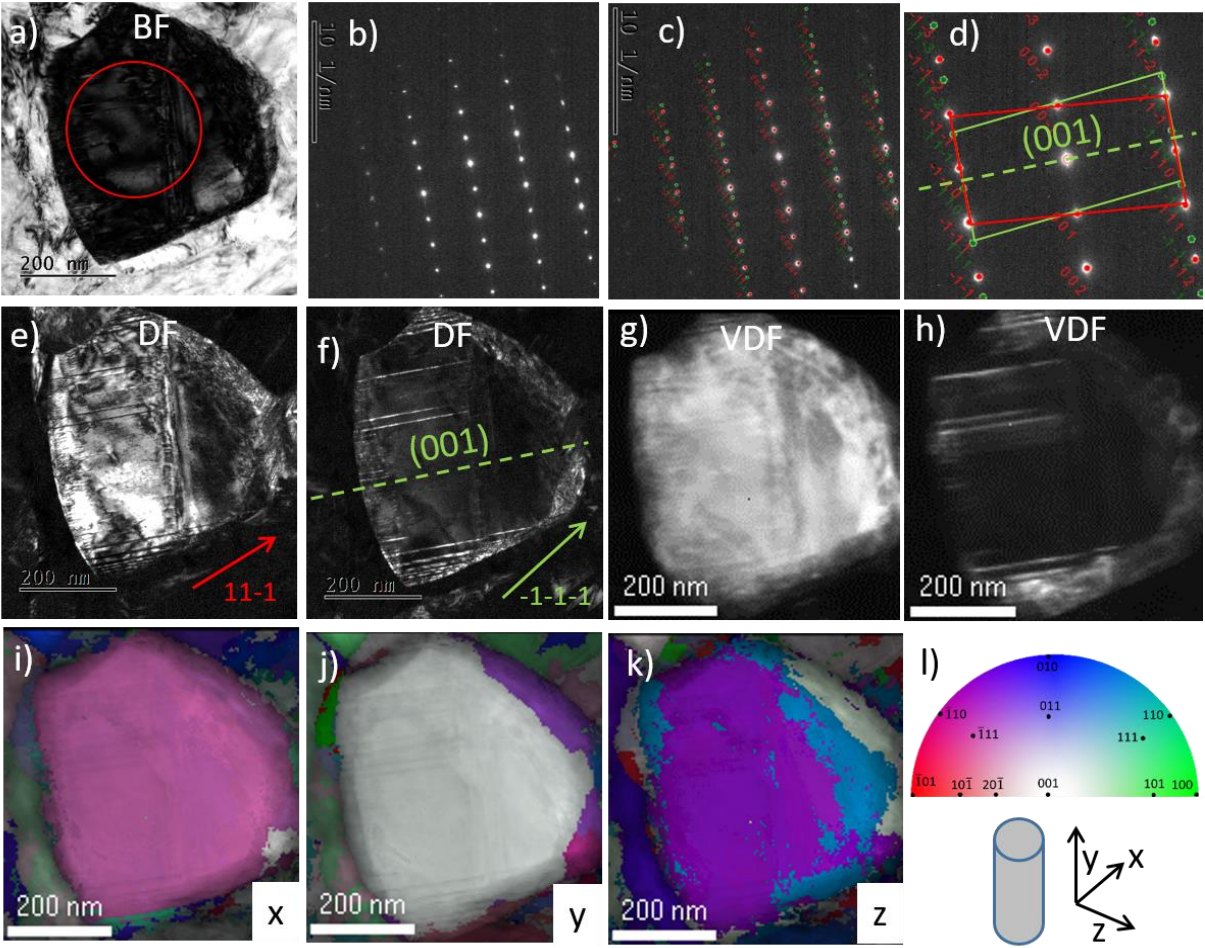

**Figure 12: Martensite variant microstructures in a partially detwinned grain of 15 ms NiTi #5 shape memory wire deformed up to end of the stress plateau 7.5% strain (Fig. 8).** a) BF image of the selected grain oriented in [1-10] zone, b,c,d) electron diffraction pattern taken from the SAED area denoted in (a) showing two diffraction patterns corresponding to B19' lattices in (001) compound twin relation, e,f) DF images of the grain taken using denoted reflections of both martensite lattices visualizing few (001) compound twins in a martensite matrix, g,h) VDF images from the analysis of the ASTAR map of the grain, i,j,k) ASTAR reconstruction of the martensite variant microstructure in the grain presented by showing crystal orientations within the grain along the x-,y- and z-axis (along the electron beam roughly perpendicular to the lamella surface). Color scale in (l) denotes orientations of basic crystal directions of the monoclinic lattice.

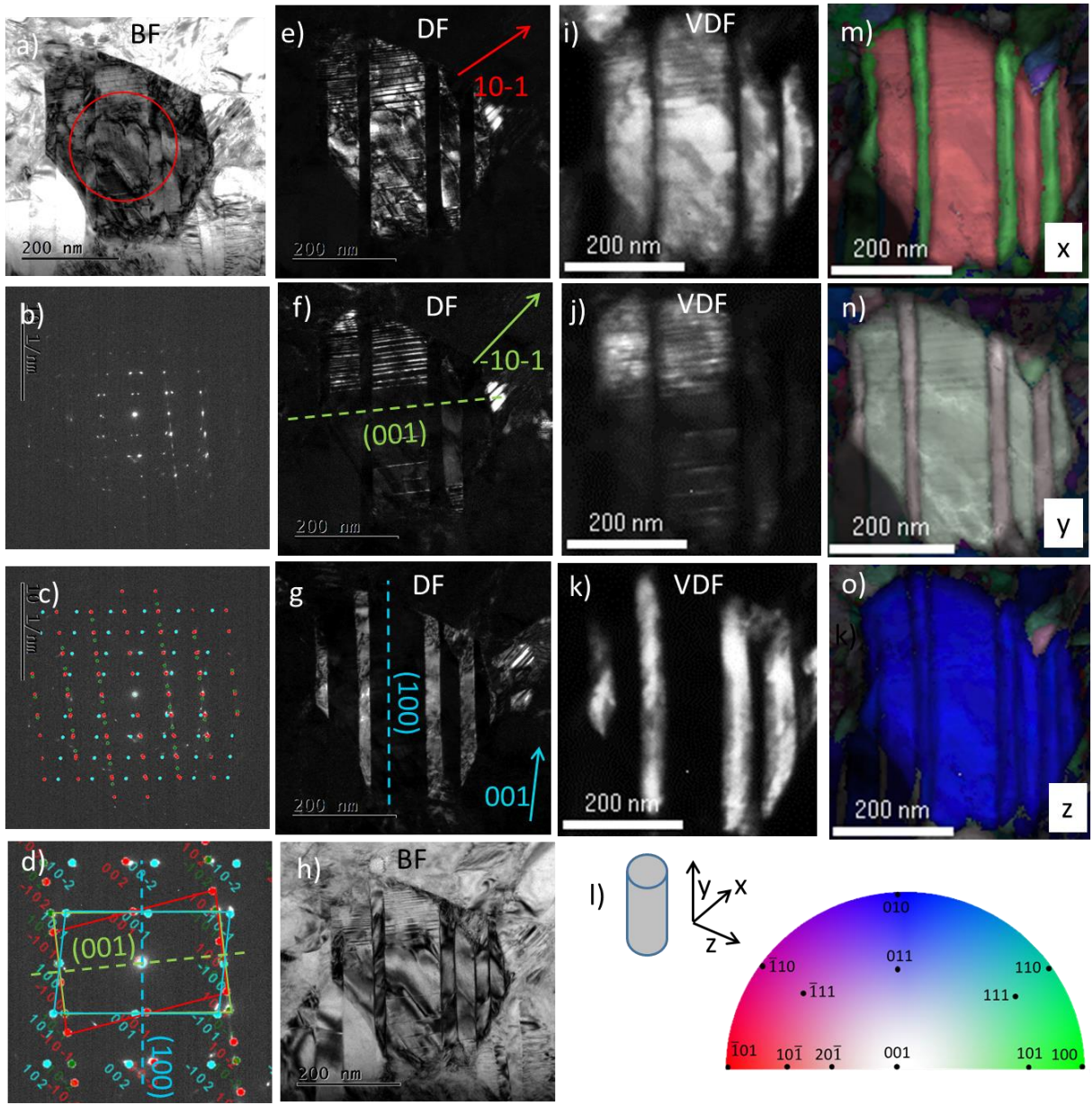

**Figure 13: Martensite variant microstructures in a partially detwinned grain with (100) twins in 15 ms NiTi #5 shape memory wire deformed up to end of the stress plateau 7.5% strain (Fig. 8).** a) BF image of the selected grain oriented in [010] zone, b,c,d) electron diffraction pattern taken from the SAED area denoted in (a) showing three diffraction patterns corresponding to martensite matrix, (001) compound twin and (100) twin , e,f,g) DF images of the grain taken using denoted reflections visualizing bright martensite matrix (e), (001) compound twins (f) and (100) twins (g), h) BF TEM image of the grain, i,j,k) VDF images from the analysis of the ASTAR map of the grain, l) color scale denotes orientations of basic crystal directions of the monoclinic lattice, m,n,o) ASTAR reconstruction of the martensite variant microstructure in the grain presented by showing crystal orientations within the grain along the x-,y- and z-axis (along the electron beam roughly perpendicular to the lamella surface).

The observation of the stress induced martensite in a form partially detwinned (001) compound twin laminates filling whole grains means that the polycrystal grains become in oriented martensite a kind of martensite crystals with a single orientation of the (001) crystal plane. This left many questions open. Why there are (001) compound twins, why these twins do not detwin, why there are no other twins and how such martensitic polycrystal deforms upon further tensile loading beyond the end of the reorientation plateau? The absence of detwinning was rationalized in [12] by considering lateral constraint of neighboring grains in highly textured NiTi polycrystalline wire (see appendix C in [12]). How the wire deformed plastically in the absence of kwinking (there are no kwink bands in the microstructure of the wire deformed up to the end of plateau (Figs. 11-13)), however, remained unclear.

No information concerning the mechanism by which the 3% plastic deformation was generated can be inferred from Figs. 11-13. The (001) compound twins and (100) twins are transformation twins which retransform back to the parent phase on heating and thus do not give rise to any plastic strains. There is dislocation like contrast in the microstructure but individual slip dislocations in the martensite matrix cannot be analyzed by conventional TEM analysis.

The observation of (100) twins in martensite variant microstructures of stress induced martensite (Fig. 13) is relatively rare. However, the (100) twins become abundant in the microstructure of the wires deformed further up to 12% (Fig. 14) and 15% (Fig. 15) strains suggesting their important role in the plastic deformation of the martensite by kwinking [13,16].

The wire deformed further beyond the end of the stress plateau up to 12% (Fig. 14) and 15% (Fig. 15) strain contained both (001) compound twinned and detwinned grains which contained (100) twin and (20-1) twin bands. While the grains in the wire deformed up to 12% (Fig. 14) contained mainly the (100) twin bands, grains in the wire deformed up to 15% (Fig. 15) were dominated by (20-1) twin bands.

The martensite variant microstructures in Figs. 14,15 can be compared with microstructures in the same wire deformed plastically up to 15% strain in the martensite state at 20 °C (Fig. 7-9 in [13]). The character and density of kwink bands is similar in both microstructures. However, there seems to be more (100) twins in the wire deformed by stress induced MT at 120 °C, particularly in Fig. 14. This is rationalized by considering that the (100) twinning always precedes the (20-1) kwinking in the plastic deformation of martensite by kwinking. While the (100) twins retransform upon unloading and heating to the parent austenite, the (20-1) kwinks retransform to {114} twinned austenite giving rise to large plastic strains (Fig. 11 in [12]).

Based on the results of our previous work on plastically deformed martensite in this wire [12,13], we know that the microstructures in Figs. 14,15 were created by plastic deformation of the B19' martensite by kwinking [13,16] taking place via [100](001) dislocation based kinking combined with (100) deformation twinning. This means that the stress induced martensite coexisting with the austenite in grains of the NiTi wire deforms plastically at stresses, which are significantly lower than the yield stress of martensitic wire. The experimentally observed low slope of the stress-strain curve at 120 °C beyond the end of the stress plateau in tensile tests at elevated temperatures (Fig. 2a, 4) is thus due to combined martensitic transformation of retained austenite and plastic deformation of stress induced martensite.

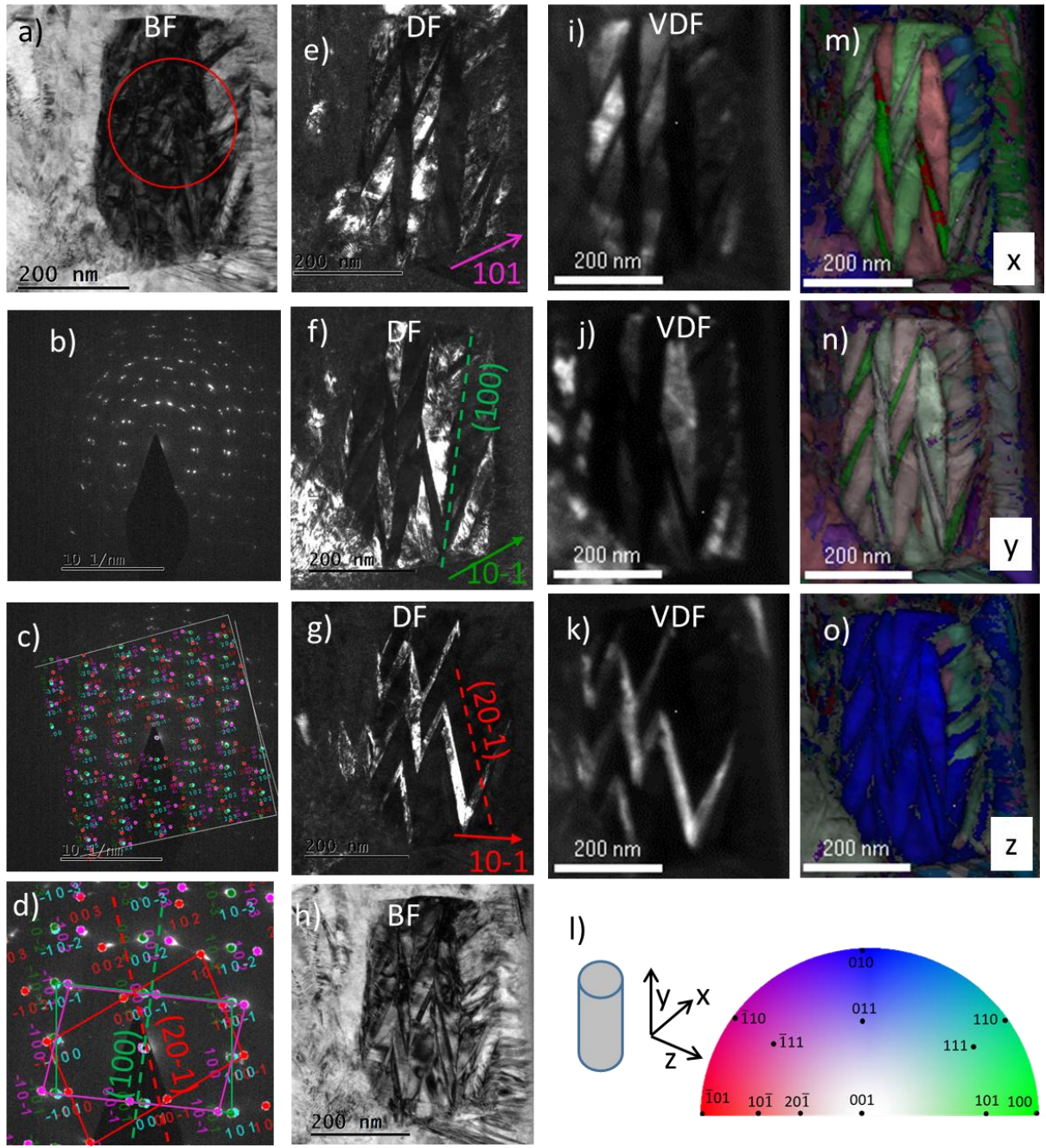

**Figure 14:** Martensite variant microstructures in a grain filled with multiple martensite variants in **15 ms NiTi #5** shape memory wire deformed up 12% strain (**Fig. 8).** a) BF image of the selected grain oriented in [010] zone, b,c,d) electron diffraction pattern taken from the SAED area denoted in (a) showing three diffraction patterns corresponding to martensite matrix, (100) twins and (20-1) kwinks created by plastic deformation of martensite by kwinking, e,f,g) DF images of the grain taken using denoted reflections visualizing bright martensite matrix (e), (100) twins (f) and (20-1) kwinks (g) nucleating from both matrix and (100) twins, h) BF TEM image of the grain, i,j,k) VDF images from the analysis of the ASTAR map of the grain, l) color scale denotes orientations of basic crystal directions of the monoclinic lattice, m,n,o) ASTAR reconstruction of the martensite variant microstructure in the grain presented by showing crystal orientations within the grain along the x-,y- and z-axis (electron beam).

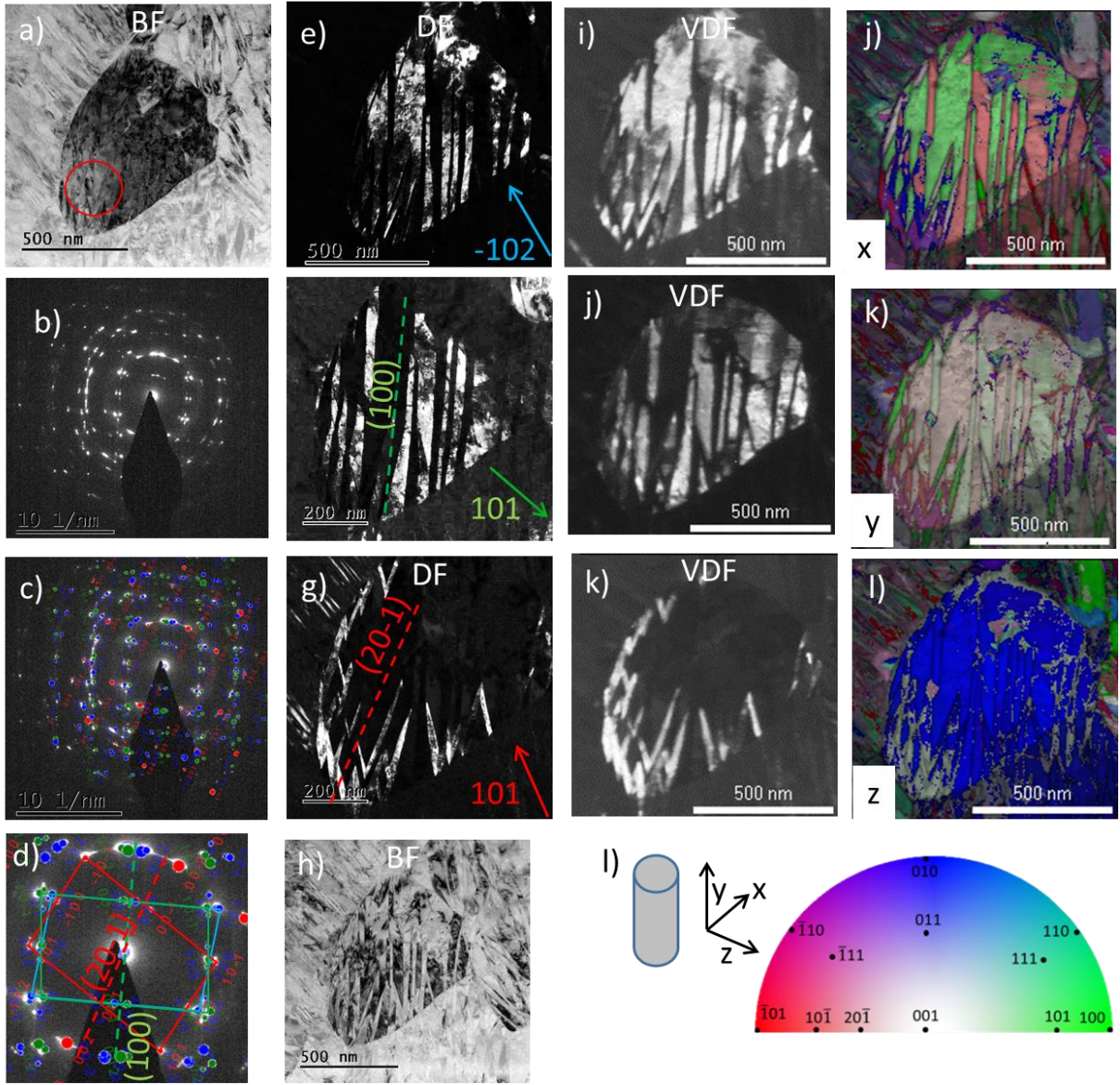

**Figure 15: Martensite variant microstructures in a grain filled with multiple martensite variants in 15 ms NiTi #5 shape memory wire deformed up 15% strain (Fig. 8).** a) BF image of the selected grain oriented in [010] zone, b,c,d) electron diffraction pattern taken from the SAED area denoted in (a) showing three diffraction patterns corresponding to martensite matrix, (100) twins and (20-1) kwinks created by plastic deformation of martensite by kwinking, e,f,g) DF images of the grain taken using denoted reflections visualizing bright martensite matrix (e), (100) twins (f) and (20-1) kwinks (g) nucleating from both matrix and (100) twins, h) BF TEM image of the grain, i,j,k) VDF images from the analysis of the ASTAR map of the grain, l) color scale denotes orientations of basic crystal directions of the monoclinic lattice, m,n,o) ASTAR reconstruction of the martensite variant microstructure in the grain presented by showing crystal orientations within the grain along the x-,y- and z-axis (electron beam). The color scale in (p) denotes orientations of basic crystal directions of the monoclinic lattice.

The observation of kwink bands in the microstructure of the wire deformed beyond the end of the stress plateau in tensile test at 120 °C (Fig. 14,15) implies that the oriented stress induced martensite deformed plastically when the wire was deformed beyond the end of the stress plateau. The plastic deformation, however, took place at stresses ~750 MPa (Fig. 8), which are lower than the yield stress for plastic deformation of martensite by kwinking ~850 MPa (Fig. 2e). This means that the ongoing martensitic transformation assisted plastic deformation of the stress induced martensite.
21

# 4. Discussion

The results of thermomechanical loading experiments on NiTi SME wire presented in section 3 suggest that the activated deformation mechanism changes with increasing test temperatures from the stress induced MT towards the coupled MT with plastic deformation and that the recoverable stress-strain-temperature response of the wire in closed loop thermomechanical load cycles (Fig. 3) becomes gradually irrecoverable. Such loss of functional properties with increasing temperature is observed in most of NiTi and NiTiX alloys. The range of temperatures at which this loss is observed depends on the chemical composition and virgin austenitic microstructure of the wire [4,20]. Functional properties of commercial strengthened NiTi alloys are lost when MTs take place in the temperature range 100-150 °C, but solution annealed NiTi alloys and some ternary alloys display such loss of functionality at significantly lower temperatures below the room temperature.

Based on the results reported in sections 3.1 and 3.2, the gradual change of the deformation mechanism with increasing temperature is evidenced by: i) stress induced MT in the plateau range becoming incomplete (Fig. 7), ii) decreasing slope of stress-strain curve beyond the end of the plateau (Fig. 4a), iii) temperature dependence of plateau stress deviating from linearity (Fig. 4a), and iv) recoverable strain evaluated within the stress plateau range decreasing with increasing test temperature (Fig. 3c). The higher the temperature at which the stress induced MT occurs, the more plastic deformation accompanies the stress induced MT within the plateau range. The experiments provide clear evidence that the stress induced MT continues upon tensile loading beyond the end of the plateau in tensile test at 120 °C (Figs. 8,9) and that the stress induced martensite deforms plastically (Figs. 14,15).

Before we start to discuss this change of deformation mechanism with increasing temperature, let us briefly mention few important experimental observations. First is that plastic strains are generated during the thermomechanical loading test only when the stress induced MT proceeds - i.e. when the stress-temperature paths cross the blue and magenta lines in the σ-T diagram (Fig. 2e) in denoted directions. Partially, this can be ascribed to the tensile deformation proceeding in a localized manner via motion of cone shaped Luders band fronts. It was reported [33] that local shear stresses within grains temporarily increase when the Luders band front passes through them. They become ~20% higher than the stresses with the austenite grains in the yet untransformed part of the wire [33]. The stress in a grain increases when it transforms to martensite within the propagating Luders band front also simply because the cross section of the wire decreases. However, the localized deformation is surely not the only reason why plastic deformation accompanies the forward stress induced MT. Main reason to be discussed below is the plastic deformation of the stress induced martensite, which cannot withstand the stress, at which it was created. Other potential reasons include: i) the requirement for strain compatibility at mobile habit plane interfaces and stationary grain boundaries and ii) load transfer from austenite to martensite in the polycrystalline environment.

The forward MT generates plastic strain at stresses, which are significantly lower than critical stresses for plastic deformation of both austenite and martensite (Figs. 2a, 5, 6, S2). Such plastic deformation assisted by the martensitic transformation, which was observed frequently in steels, has been termed transformation induced plasticity (TRIP) in the literature [39]. Although generation of plastic strains by martensitic transformations in NiTi formally fulfils this definition, it shall not be called TRIP because the mechanism is different. Both austenite and martensite deform plastically in TRIP steels and there is no strain reversibility on heating. In a contrast, stress-strain-temperature behavior of thermomechanically loaded NiTi is inherently reversible, but incremental plastic strains are generated whenever the forward and/or reverse MTs occur under stress in cyclic thermomechanical loads [3]. Plastic strains are generated not only during forward MT on loading while the stress increases, but also during the reverse MT on unloading while the stress decreases [3,38]. In this work, we focus on the forward MT. Reverse MT proceeding in all experiments upon stress free heating (Figs. 5-9) does not generate plastic strains [2].

**4.1 Recovered and unrecovered strains generated by the forward martensitic transformation**

As the test temperature increases in the range 70 to 150 °C (Figs. 2a,3,6), the deformation mechanism gradually changes, tensile strain becomes unrecoverable and functional behavior of the NiTi wire is completely lost at 150 °C. As the plateau stress, at which the stress induced MT proceeds, increases from 100 to 600 MPa with increasing temperature, the plastic strain generated by the forward MT increases, but recoverable strain remains approximately constant (Figs. 6,7).

The recoverable strain actually increases from zero to 5% strain in the temperature range 65-80 °C close above $M_S$ temperature. The forward MT in this temperature range was investigated in experiments involving cooling the 15 ms NiTi #5 SME wire under various constant stresses [18], where information on martensite variant microstructures and martensite textures created by forward MT upon cooling under stress can be obtained.

A question is why the maximum recoverable transformation strain reaches only ~5% (Figs. 7,9), if the B2-B19' transformation in <111>A fiber textured NiTi wire can provide up to 9 % recoverable transformation strain (Fig. S1). We explain this by the observation of (001) compound twin laminates in the microstructure of NiTi wires transforming under high stress 500 MPa (Figs. 11-13). We assume that the (001) compound twins were introduced by the habit plane of the stress induced MT, as will be discussed in chapter 4.3.

Main difference between the superelastic NiTi #1 NiTi wires and the NiTi #5 SME wire used in this work, besides the difference in transformation temperatures, consists in the lower resistance of the present NiTi #5 SME wire to the plastic deformation accompanying the martensitic transformation. Once the stress induced martensite starts to deform plastically in the tensile test, unrecovered plastic strain and cooling strain sharply increase with the increasing temperature (Fig. 6,7). We explain this by the plastic deformation of martensite, which causes the martensitic transformation in the plateau range to become incomplete with increasing temperature. Indirect evidence for the incomplete martensitic transformation is provided by the stress-strain-temperature responses in Fig. 6 (strain can increase on cooling only if there is austenite in the wire). Direct experimental evidence on the incomplete MTs in the plateau range of superelastic tensile test was obtained by in-situ synchrotron x-ray diffraction [37].

Results of the experiments in Figs. 8,9 focusing stress induced MT in the tensile test at ~120 °C show that the unrecovered plastic strain increases while the recovered strain remains approximately constant when the wire is loaded beyond the end of the plateau. The decreasing cooling strain with increasing stress (Fig. 9b) proves that the volume of the retained austenite existing in the wire loaded up to the end of the plateau decreases upon further loading beyond the end of the plateau. This means that the retained austenite in the NiTi wire deformed beyond the end of the plateau continues to transform, but the martensitic transformation gives rise mainly to plastic strains. In summary, the slope of the stress-strain curve beyond the end of the plateau decreasing with increasing test temperature (Fig. 4a), suggest that the NiTi wire deforms via martensitic transformation coupled with plastic deformation.

**4.2 Martensite variant microstructure created by the stress induced martensitic transformation**

The analysis of martensite variant microstructures in grains of the NiTi wire deformed up to the end of the stress plateau at 120 °C showed that the stress induced martensite is mainly formed by single domain (001) compound twin laminates in grains (with different level of detwinning and occasional (100) twins). This is similar to martensite variant microstructures observed in the same NiTi wire deformed up to the end of the stress plateau at 100 °C [12]. The only difference is that plastic strains and volume fraction of retained austenite generated by the forward MT at 120 °C are significantly larger. It comes out from Fig. 4, that smaller plastic strains and less retained austenite were also in the martensite variant microstructure at 100 °C [12], we just did not know about them. This is because no information on plastic deformation in martensite can be obtained by analysing martensite variant microstructures in Figs. 11-13. Nevertheless, since nearly half of the plateau strain corresponds to unrecoverable plastic strain, plastic deformation must have accompanied the forward MT during the tensile deformation within the plateau range.

**4.3 Strain compatibility at habit plane interfaces and grain boundaries**

According to the widespread view in the literature, stress induced MT in NiTi proceeds into <011> type-II twinned martensite [5,7,8,29], as predicted by the PTMC theory [9]. However, since no <011> type-II twin laminates were observed in the analyzed martensite microstructures (Figs. 11-13), stress induced martensitic transformation involving habit plane interfaces between austenite and type-II twinned martensite can be excluded. Instead, most of the grains of the NiTi wire deformed up to the end of the stress plateau at 120 °C (Figs. 11-13) are filled with partially detwinned single domain laminates of (001) compound twins. Although the (001) compound twinning cannot serve as LIS in the B2-B19' transformation [1], the forward stress induced MT proceeds into (001) compound twinned martensite.

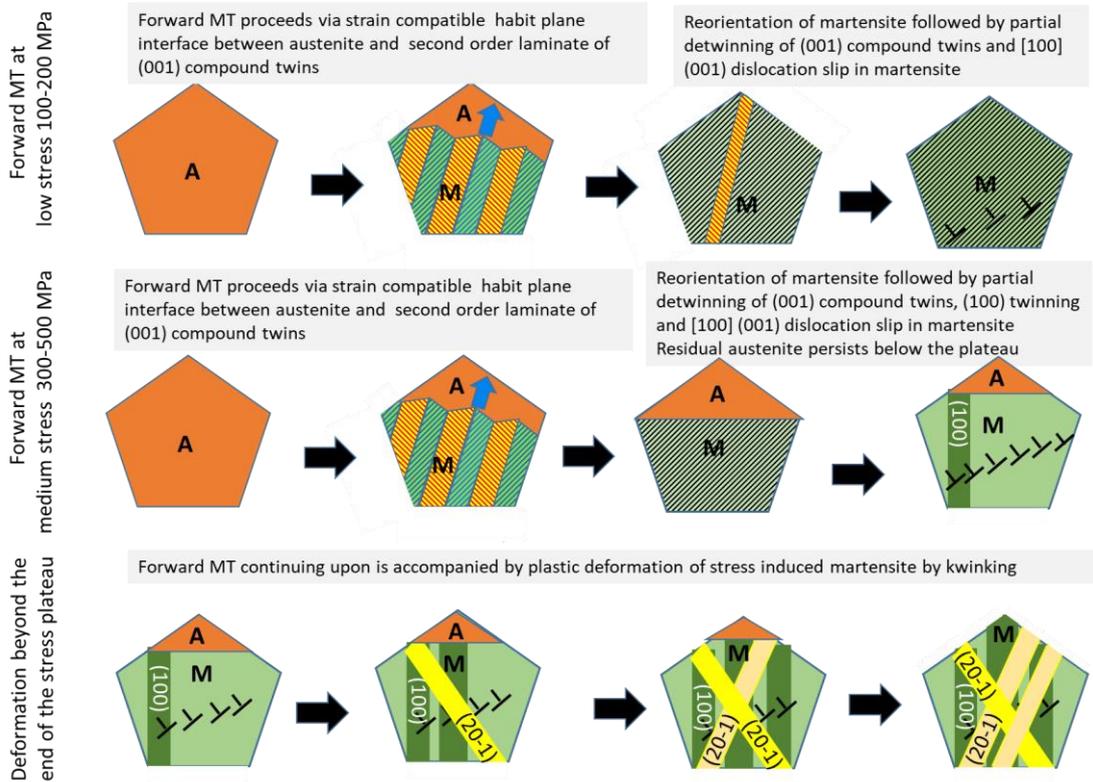

**Figure 16: Schematic figure suggesting habit plane interface between austenite and martensite established during forward stress induced martensitic transformation** within the transformation plateau range at low 100-200 MPa stresses (upper row) and medium 300-500 MPa stresses (mid row). The bottom row characterizes evolution of microstructure in the wire deformed 120 °C (500 MPa) beyond the end of the stress plateau.

Since the strain compatibility at the habit plane cannot be achieved via habit plane between austenite and single (001) compound twin laminate, it must be achieved differently. As we have no experimental evidence on the habit planes of the forward MT, we simply assume strain compatible interface between the austenite and second order (001) compound twin laminate in martensite, as originally proposed by Waitz [41] for thermally induced martensite in nanocrystalline NiTi. In case of martensitic transformation under external stress, however, we simply assume the stress induced martensite immediately reorients into single domain occupying whole grain (Fig. 11). Simultaneously, some grains detwin (Fig. 12), some grains undergo (100) twinning in martensite (Fig. 13) and the martensite deforms via [100](001) dislocation slip (section 4.4). In addition, there is the retained austenite at the end of the stress plateau, the volume fraction of which increases with increasing temperature. Habit plane of the stress induced martensitic transformation in tensile test at 120 °C is schematically shown in the middle row in Fig. 16. We understand such stress induced forward MT as a single event occurring within the martensite band front propagating through the NiTi wire in the plateau range of the tensile test at stresses ~500 MPa.

Assuming such habit planes of the stress induced MT, we can explain the presence of (001) compound twins in the microstructure of stress induced martensite (Figs 11-13), but other questions remain. Why the (001) compound twins do not detwin was already explained by considering the lateral constraint of neighboring grains in highly textured NiTi polycrystalline wire (see appendix C in [12]). Nevertheless, stress induced MT into more and more detwinned grains gives rise to large shape strains that are mutually incompatible within the transforming polycrystalline wire. This strain incompatibility represents additional constraint, which prevents completion of the MT within the transformation plateau range and forces the martensite to deform plastically. This explains why the forward stress induced MT in tensile tests at 120 °C generates plastic strains (Fig. 7).

The upper plateau strains evaluated in tensile tests at first increase and later decrease with increasing temperature (Figs. 6,7). To explain that, it is assumed that, as the temperature increases, plastic deformation of martensite by dislocation slip becomes more and more pronounced. Simultaneously, since deformation of neighboring grains via stress induced MT is not compatible at grain boundaries, the stress induced MT gradually becomes incomplete within the plateau range (completion requires stress increase). The higher the plateau stress, the larger volume fraction of retained austenite (Fig. 7). The upper plateau strain decreases at highest temperatures due to the increasing amount of the retained austenite. Simultaneously, the plastic strain generated by the forward MT under stress increases while the recoverable transformation strain remains constant with increasing temperature (Fig. 7).

There is also a theoretical possibility the stress induced MT proceeds at high stresses via habit plane between austenite and single variant of martensite [25], while the strain compatibility at the habit plane is facilitated elastic [26] and/or plastic [24] deformation. Nevertheless, since the (001) compound twins were observed not only in the microstructure created by forward stress induced martensitic transformation in tensile test at 120 °C within the stress plateau at 500 MPa (Figs. 11-13), but also in the martensite induced by cooling from austenite under 500 MPa stress [18], the most plausible explanation is that the stress induced MT in nanocrystalline NiTi proceeds via habit plane between austenite and second order (001) compound twin laminate followed by martensite reorientation even at high stresses 500-700 MPa (Fig. 2d).

**4.4 Plastic deformation of the stress induced martensite via dislocation slip**

Assuming that plastic deformation accompanying the forward martensitic transformation under stress occurs in the martensite but not in the austenite, we have to specify how the stress induced martensite deforms plastically. As was already mentioned in the introduction, only two mechanisms of plastic deformation of the stress induced B19' martensite were so far considered in the SMA literature: i) [100](001) dislocation slip [13,18,28] theoretically predicted to be activated at low stresses and ii) kwinking deformation [13,16] experimentally confirmed to be activated at high stresses. Since we did not found any signs of kwink bands in martensite variant microstructure created by forward stress induced MT in tensile test at 120 °C (Figs. 11-13), we assume that the experimentally observed plastic strains occurred via dislocation slip in the stress induced martensite phase.

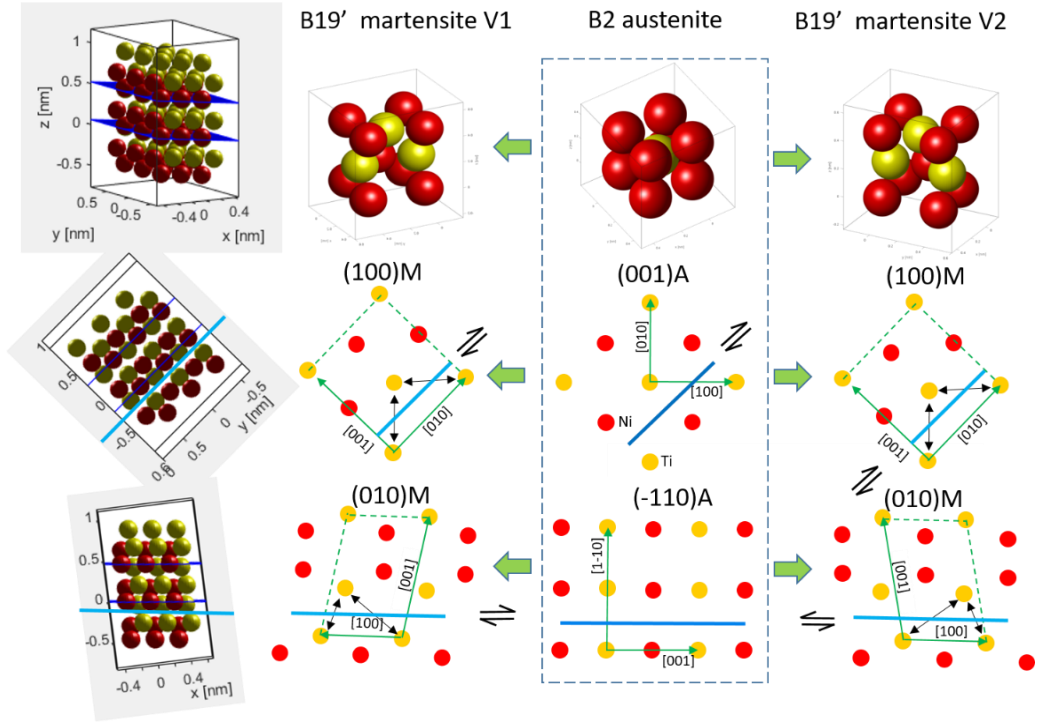

**Figure 17: Dislocation slip in the monoclinic B19'structure.** Crystal structures of B2 austenite and B19' martensite are visualized by cell models built using the CrystBox tool [42]. The sketch shows how the crystal structure change during the the B2-B19' martensitic transformation by showing atom positions on (001)A and (110)A cross sections along austenite planes changing into their positions on (100)M and (010)M cross sections along martensite planes, respectively. Two layers of atoms are plot together without differentiating between them. There are week Ti-Ti bonds (denoted by double black arrows) allowing for easy shearing the monoclinic structure [43]. Shearing in the [100] direction enables (001) compound twinning (lattice correspondence variants V1 and V2 can be converted one into another by the shear applied on (001) plane along the +/- [100] shear direction) and dislocation slip in [100](001) direction. The [100](001) and [010](001) martensite slip systems are inherited into austenite as <100>{110} and <110>{110} slip systems, respectively.

In order to understand why the [100](001) dislocation slip becomes activated in stress induced martensite, we need to delve into the crystallography of the B2-B19' transformation in NiTi. Fig. 17 visualizes atomic structures of B2 austenite and B19' martensite as "cells" built by the software CrysTBox [44] utilized in the TEM analysis of martensite variant microstructure [14]. Lattice parameters $a_0$ of the B2 austenite and $a < b < c$ and $\beta$ monoclinic angle of the B19' martensite are given in Table 1. The shape strain associated with the B2-B19' martensitic transformation involves expansion along the c-axis direction and contraction around the a- and b- axis directions, the $\beta$ angle changes from 90° to 96.8° via shear on the (001) plane in the [100] direction [1, 15]. Moreover, selected atoms slightly change their positions within the unit cell (shuffle). 2D projections of both lattices in the (001)A and (1-10)A austenite planes and correspondent (100)M and (010)M martensite planes visualize this distortion.

The low symmetry monoclinic B19' structure can deform plastically only via dislocation slip on the (001) plane (a single most densely packed lattice plane of the B19' structure with largest lattice spacing).

Deformation in other slip systems is energetically much more expensive [28]. Therefore, the monoclinic B19' structure displays extreme anisotropy of plastic slip. Besides of that, the monoclinic structure can be easily sheared on the (001) plane in the [100] direction because of the weak Ti-Ti bonds [43] (indicated by double arrows in Fig. 17). This easy shearing enables (001) compound twinning between the martensite variants V1 and V2 (Fig. 18) which facilitates detwinning of (001) compound twin laminates in grains taking place along with the [100](001) dislocation slip.

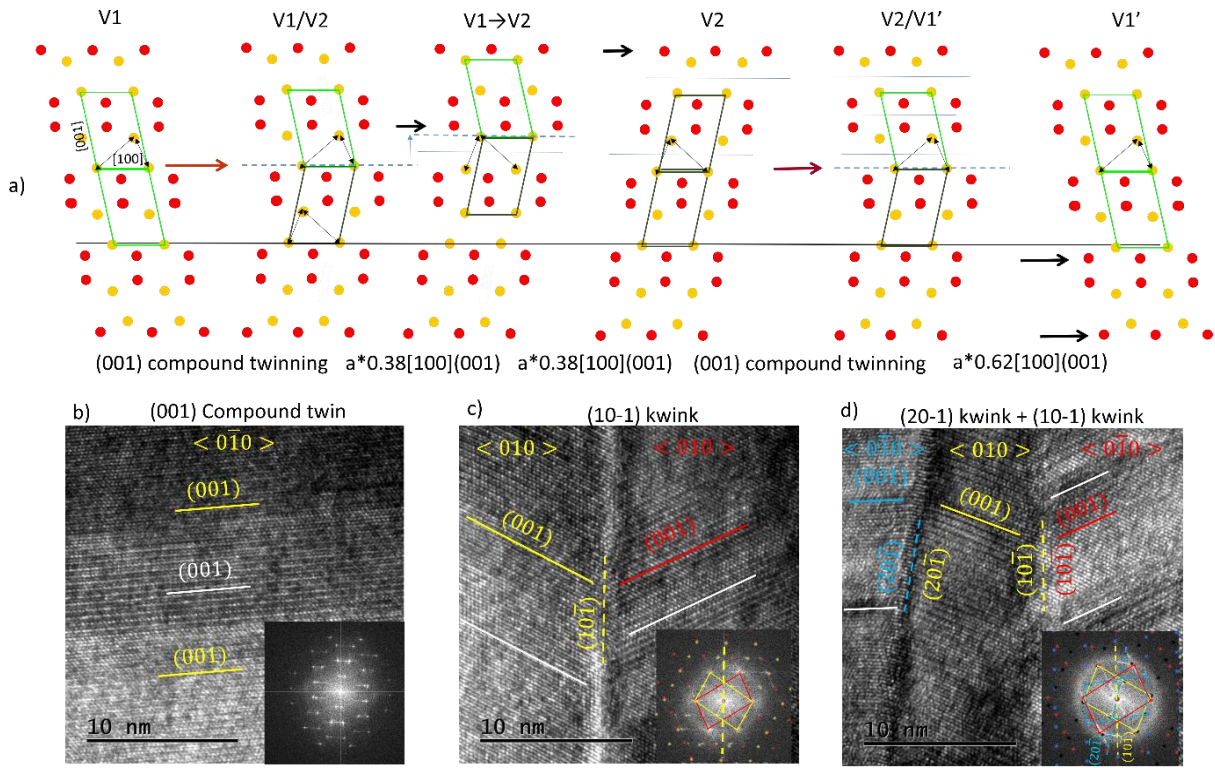

**Figure 18: Plastic deformation of the B19'martensite via (001) compound twinning and [100](001) dislocation slip.** Plastic deformation of the martensite variant V1 by the a*[100](001) dislocation slip (from V1 to V1') can be decomposed into (001) compound twinning (from V1 to V2) followed by plastic deformation of the variant V2 by coordinated slip of partial dislocation a*0.62[100](001) (from V2 to V1'). The (001) compound twinning can be imagined as proceeding via coordinated slip of partial dislocations a*0.38 [100](001) and/or a*0.62[100](001) on the (001) crystal plane. HRTEM images showing b) (001) compound twin in martensite, c) (10-1) kwink and (20-1) kwink and (10-1) kwink in plastically deformed martensite document plastic deformation of martensite via plastic slips on (001) plane. See Refs [13,16] for more information on plastic deformation of B19' martensite via kwinking.

The critical stresses for both (001) compound twinning as well as dislocation slip are low [28]. In spite of that, the (001) compound twin laminates were observed in grains of the NiTi wire deformed up to the end of the stress plateau at 500 MPa stress (Figs. 11-13). This was explained in Ref. [12] by assuming that the (001) compound twin laminates in grains cannot detwin due to the constraint from neighboring grains in the deformed NiTi polycrystal (for detailed explanation see Appendix C in [12]).

However, plastic deformation via [100](001) dislocation slip is restricted via the same polycrystalline constraint, since it provides similar strains as the detwinning of (001) compound twin laminates (slip plane and twinning plane as well as Burgers vector and twinning shear direction are parallel (Fig. 17)).

The difference is that strains due to dislocation slip are not recoverable and are not limited by the crystallography of the MT. The plastic deformation of martensite via dislocation slip in [100](001) plane occurs regardless the stress induced martensite is (001) compound twinned or not.

Although the dislocation slip in (001) crystal plane of martensite is effectively suppressed in many NiTi wires (Fig. S1), it is not the case for the 15 ms NiTi #5 wire (Fig. S2), since small plastic strains are generated already by the reorientation process at 150 MPa (Figs. S3-S6). Therefore, it is not surprising that the forward MT upon cooling under stresses around 200 MPa (Figs. 6,7) generates small plastic strains as well. It is assumed that the [100](001) dislocation slip competes with detwinning of (001) compound twins (Fig. 18) during the forward stress induced MT within the plateau range. The grains within the NiTi wire cooled under stress are assumed to adjust their shape in martensite state by combination of elastic deformation, detwinning of (001) compound twin laminates and [100](001) dislocation slip in martensite.

### 4.5 Plastic deformation of stress induced martensite via kwinking

It was concluded in the section 3.2 (Figs. 8,9) that the retained austenite at the end of the stress plateau at 120 °C gradually transformed to the martensite when the wire was deformed further beyond the end of the stress plateau. Simultaneously, as the wire was deformed beyond the end of the stress plateau, the reconstructed martensite variant microstructures (Figs. 14,15) contained increasing density of (20-1) kwink bands. This provides clear evidence that the stress induced martensite deformed plastically via kwinking deformation mechanism at stresses lower than the yield stress $\sigma^Y_{M\_kwink}$ (Fig. 2e).

The sequence of reconstructed martensite variant microstructures observed in NiTi wire deformed beyond the end of the stress plateau at 120 °C (Figs. 11-15) in fact very nicely documents the kwinking deformation mechanism [13,16]. When the wire is deformed beyond the end of the plateau, a grain filled with partially detwinned (001) compound twin laminates (Figs. 11,12) undergoes (100) deformation twinning (Fig. 13). As the stress further increases, the (20-1) kwink bands appearing within the matrix as well as within the (100) twins form wedges (Fig. 14). Further (100) and (20-1) kwinking gives rise to wedge microstructures (Fig. 15) characteristic for plastically deformed B19' martensite [13,14].

### 4.5 Loss of functional behavior with increasing test temperature

Finally, Fig. 19 offers a kind of summary on the loss of functional behavior of NiTi with increasing test temperature. As the test temperature increases from 80 °C to 150 °C (Fig. 19a), the plastic strain generated by the stress induced MT within the stress plateau (Fig. 7a) as well as the volume fraction of the retained austenite at the end of the stress plateau gradually increase (evidenced by the increasing cooling strain within creasing temperature (Figs. 7a,19c). The deformation mechanism acting during tensile tests thus gradually changes with increasing test temperature from the conventional stress induced MT at 80 °C towards the coupled MT with plastic deformation at 150 °C. The change of the deformation mechanism resulting in sharp loss of strain recoverability (Fig. 3) is best evidenced by the decreasing slope of the stress-strain curve beyond the end of the plateau with increasing test temperatures and increasing deviation of the upper plateau stress from the linear temperature dependence observed in tensile tests until fracture (Fig. 19c).

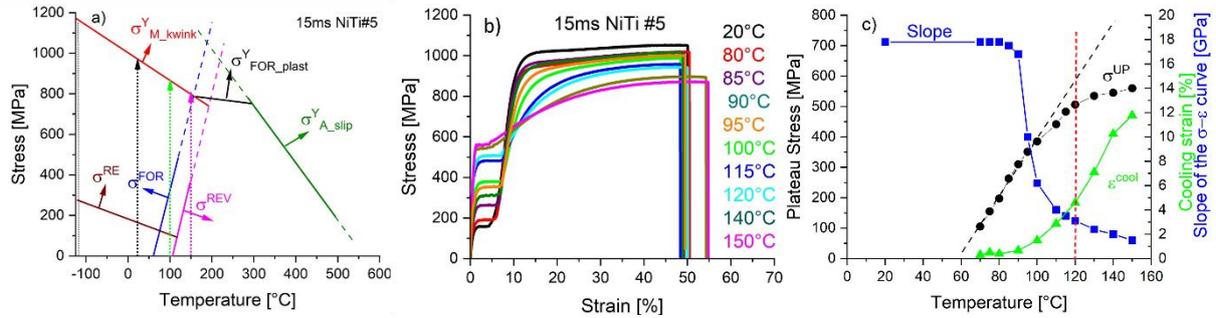

**Figure 19: Stress induced martensitic transformation in tensile tests on 15 ms NiTi #5 wire at increasing test temperatures.** a) σ-T diagram, b) stress-strain curves in tensile tests until fracture at increasing temperatures, c) temperature dependence of transformation plateau stress, slope of the stress-strain curves beyond the end of the plateau and cooling strain from the experiment in Figs. 6,7. Deashed line in (c) denotes the temperature 120 °C at which the stress induced MT was investigated in detail (Figs. 8-15).

As concerns the recoverable transformation strain, one cannot rely of the plateau length, it increases, reaches a maximum and decreases with increasing test temperature (Figs. 3d,6a). Similar results were observed on superelastic NiTi wire [34, 37]. The length of the plateau depends on transformation strain as well as on the plastic strain. It was found that the volume fraction of martensite at the end of the stress plateau could be as low as 10% in tensile tests at high temperatures (i.e. 90% retained austenite) [37]. This retained austenite gradually transforms to martensite during the tensile test but hardly ever reaches 100% at the fracture, Simultaneously, the activity of kwinking deformation is reduced at highest test temperatures [34]. The same is expected to be the case in present 15 ms NiTi #5 SME wire when it is deformed at high temperatures as e.g. 150 °C (Fig. 2d). The wire deforms via coupled MT with plastic deformation in the temperature range from ~150 °C up to ~300 °C (Fig. 2b,e). This deformation mechanism is, however, far from being understood and needs to be further investigated.

Finally, we would like to make two comments. First is that functional properties of NiTi are lost with increasing temperature not only due to plastic strains accompanying the forward MT but also due to plastic strains generated by the reverse MT [3]. This was so far neglected for simplicity. Second, the NiTi wires can be strengthened by cold work/annealing at lower temperatures or by long time aging at low temperatures which both suppress the [100](001) dislocation slip in martensite. For example, 10 ms NiTi #5 SME wire strengthened by cold work/annealing displays reversible superelastic stress-strain response at 120 °C [4]. Nevertheless, the mechanism of the degradation of functional behavior with increasing temperature remains the same. Only the incremental plastic strains become smaller and degradation of functional behavior is shifted to higher temperatures.

# 5. Conclusions

Stress induced forward B2-B19' martensitic transformation in 15 ms NiTi#5 shape memory wire was investigated by thermomechanical testing. Recoverable transformation strains and plastic strains generated by the forward martensitic transformation proceeding at various stresses were evaluated. Martensite variant microstructures created by tensile deformation at 120 °C up to various maximum strains were reconstructed by post-mortem analysis in TEM. Following conclusions were obtained:

1. The forward martensitic transformation in tensile tests on 15 ms NiTi#5 SME wire at elevated temperatures generates both recoverable transformation strain and unrecoverable plastic strain. The transformation is not completed at the end of the stress plateau. The higher the tensile test temperature, the larger the volume fraction of retained austenite at the end of the stress plateau and the magnitude of the generated plastic strain.
2. The recoverable transformation strain generated by the forward martensitic transformation within the stress plateau increases with increasing test temperature, reaches ~6% strain at 80 °C (200 MPa plateau stress) and stays nearly constant with further increasing the test temperature (plateau stress).
3. Martensite variant microstructures in grains of the wire deformed in tensile test at 120 °C up to the end of the stress plateau (7.5% strain, 700 MPa stress) consist of single domain (001) compound twin laminates, some grains were detwinned, some contained (100) twin bands.
4. The forward stress induced martensitic transformation proceeds via habit plane between austenite and second order laminate of (001) compound twins, the martensite immediately reorients and deforms plastically via [100](001) dislocation slip in martensite at low stresses and via kwinking at high stress above 500 MPa.
5. The maximum recoverable transformation strain (~6%) is smaller than the theoretically achievable strain (11%) because the stress induced martensite is (001) compound twinned and because the detwinning is constrained in polycrystalline wire. The maximum recoverable transformation strain is affected by the strength of the <111> austenite fiber texture of the wire and possibly by the [100](001) dislocation slip in martensite, which substitutes detwinning of (001) compound twins.
6. Martensite variant microstructures in grains of the wire deformed in tensile test at 120 °C further beyond the end of the stress plateau (12% and 15% strain) contain (100) twin bands and (20-1) kwink bands evidencing plastic deformation of the stress induced martensite.
7. The slope of the stress-strain curve upon loading beyond the end of the stress plateau decreases with increasing test temperature because the stress-strain response in this stage is not only elastic, but the retained austenite in the wire transforms to martensite, while the martensite deforms plastically via [100](001) dislocation slip and kwinking.
8. The loss of functional properties of NiTi with increasing temperature does not originate from the onset of plastic deformation of the austenite but from the loss of the resistance of the stress induced martensite against plastic deformation via dislocation slip and kwinking under increasing stress.

## Acknowledgment

Support from Czech Science Foundation (CSF) projects 22-15763S (Heller) and 22-20181S (Sittner) is acknowledged. P. Šittner acknowledges support from Czech Academy of Sciences through Praemium Academiae. MEYS of the Czech Republic is acknowledged for the support of infrastructure projects, CNL (CzechNanoLab LM2018110) and Ferrmion (CZ.02.01.01/00/22_008/0004591).